\documentclass[twocolumn,showpacs,preprintnumbers,amsmath,amssymb,groupedaddress,prc,floatfix]{revtex4}
\usepackage{mathrsfs}
\usepackage{longtable}
\usepackage{graphicx}
\usepackage{graphics}
\usepackage{epsfig}
\usepackage{dcolumn}
\usepackage{rotating}

\begin{document}

\title{Single Particle Excitations, Band Structures and Octupole Correlation in $^{65}$Zn}

\author{Anil Sharma}
\affiliation{UGC-DAE Consortium for Scientific Research, Kolkata Centre, Kolkata 700098, India}
\author{S. Nandi}
\affiliation{UGC-DAE Consortium for Scientific Research, Kolkata Centre, Kolkata 700098, India}
\author{S. S. Dutta}
\affiliation{UGC-DAE Consortium for Scientific Research, Kolkata Centre, Kolkata 700098, India}
\author{S. Kundu}
\affiliation{UGC-DAE Consortium for Scientific Research, Kolkata Centre, Kolkata 700098, India}
\author{Pankaj K. Giri}
\altaffiliation{Present Address: Institute of Modern Physics, Chinese Academy of Sciences, Lanzhou 730000, China}
\affiliation{UGC-DAE Consortium for Scientific Research, Kolkata Centre, Kolkata 700098, India}
\author{A. Das}
\affiliation{UGC-DAE Consortium for Scientific Research, Kolkata Centre, Kolkata 700098, India}
\author{ S. S. Ghugre}
\affiliation{UGC-DAE Consortium for Scientific Research, Kolkata Centre, Kolkata 700098, India}
\author{S. S. Nayak}
\affiliation{Variable Energy Cyclotron Centre, Kolkata 700064, India}
\affiliation{Homi Bhabha National Institute, Mumbai 400094, India}
\author{S. Basu}
\affiliation{Variable Energy Cyclotron Centre, Kolkata 700064, India}
\affiliation{Homi Bhabha National Institute, Mumbai 400094, India}
\author{S. Pal}
\affiliation{Variable Energy Cyclotron Centre, Kolkata 700064, India}
\affiliation{Homi Bhabha National Institute, Mumbai 400094, India}
\author{S. Das}
\affiliation{Variable Energy Cyclotron Centre, Kolkata 700064, India}
\affiliation{Homi Bhabha National Institute, Mumbai 400094, India}
\author{S. Dar}
\affiliation{Variable Energy Cyclotron Centre, Kolkata 700064, India}
\affiliation{Homi Bhabha National Institute, Mumbai 400094, India}
\author{S. Paul}
\affiliation{Variable Energy Cyclotron Centre, Kolkata 700064, India}
\affiliation{Homi Bhabha National Institute, Mumbai 400094, India}
\author{A. Pal}
\affiliation{Variable Energy Cyclotron Centre, Kolkata 700064, India}
\affiliation{Homi Bhabha National Institute, Mumbai 400094, India}
\author{S. Basak}
\affiliation{Variable Energy Cyclotron Centre, Kolkata 700064, India}
\affiliation{Homi Bhabha National Institute, Mumbai 400094, India}
\author{G. Mukherjee}
\affiliation{Variable Energy Cyclotron Centre, Kolkata 700064, India}
\affiliation{Homi Bhabha National Institute, Mumbai 400094, India}
\author{S. Bhattacharyya}
\affiliation{Variable Energy Cyclotron Centre, Kolkata 700064, India}
\affiliation{Homi Bhabha National Institute, Mumbai 400094, India}
\author{R. Raut}
\altaffiliation{Corresponding Author: rajarshi.raut@gmail.com}
\affiliation{UGC-DAE Consortium for Scientific Research, Kolkata Centre, Kolkata 700098, India}
\date{\today}

\begin{abstract}

The excitation scheme of the $^{65}$Zn ($Z = 30, N = 35$) nucleus has been probed following its population in the $^{63}$Cu($\alpha$,pn) reaction at E$_{beam}$ = 30 MeV and using 
an array of Compton suppressed HPGe clovers as the detection system. This work has identified several new transitions of the nucleus and have modified the placements 
of some of the previously known ones. The multipolarities and the electric/ magnetic nature of the observed $\gamma$-ray rays have been measured, using the conventional methodologies.
The spin-parity assignments for the levels have consequently been made; some of the spin-parities are new while others are either validation of the existing values or are modified results 
based on the present analysis. The experimental level scheme exhibits collective as well as single particle structures. The measured level energies have been 
compared with those calculated in the framework of the large basis shell model using a model space
of $p_{3/2}, f_{5/2}, p_{1/2}, g_{9/2}$ orbitals and two different interactions. The collective excitations of the nucleus were probed through the properties of its band structures and through the 
calculations of the Total Routhian Surface (TRS) for the associated deformations/ shapes. The results of this study brings out the essential features 
of evolving structural characteristics and developing collectivity with increasing number of nucleons outside a doubly-magic core and with their occupancy of deformation driving high-$j$ orbitals. 

\end{abstract}


\maketitle

\section{Introduction}

Systematic studies of nuclear excitations, with increasing number of nucleons outside shell closures, facilitate 
significant insights on the evolving and coexisting mechanisms for generation of angular momentum in atomic nuclei.
Such endeavors in the vicinity of the doubly-magic $^{56}$Ni-core (Z, N = 28) have, for instance, led to illustrative observations.
These include both single particle excitations, interpreted in the framework of the shell model, as well as 
rotational sequences that manifest collectivity and deformation characteristics of the nuclei. 
The model space consisting of $p_{3/2}, f_{5/2}, p_{1/2}, g_{9/2}$ orbitals outside the closure at Z, N = 28 provide 
a befitting premise for probing such varied excitation phenomena that set in subject to the changing occupancies 
of the $fp$ orbitals and the deformation driving $g_{9/2}$ orbital. Indeed, the results following the 
spectroscopic studies on the isotopes of Ni (Z = 28) \cite{Sam19,Bha23}, Cu (Z = 29) \cite{Sam18,Cha23,Sar24}, Zn (Z = 30) \cite{Zha25,Aya22}, 
Ga (Z = 31) \cite{Zho24,Sha26} etc., obtained in recent years,
bring forth the many prospects of nuclear structure pursuits in the (A$\sim$60) region. \\

The Zn isotopes have, accordingly, been subjects of some of the aforementioned investigations. The most recent of these,
on $^{67}$Zn \cite{Zha25} identified collective structures in its excitation scheme including one associated with the octupole degree of freedom.
The study on $^{66}$Zn \cite{Aya22} established single particle excitations dominating the low-spin regime of the nucleus and a rotational band 
at higher excitation that is characteristically similar to the superdeformed bands previously observed in the region. 
The neutron deficient isotopes $^{63,64}$Zn have been studied \cite{Sin98,Gho19,Kar04} in multiple experiments that essentially validated
the same (aforementioned) excitation mechanisms underlying the level structure of these nuclei. As was cited by Karlgren {\it{et al.}} \cite{Kar04},
while discussing the spectroscopic findings on $^{64}$Zn, the band structures in these nuclei are of significance in underscoring 
the onset of collectivity even within the limited valence space outside the closure at N, Z = 28. The most recent 
investigation on the level structure of $^{63}$Zn \cite{Gho19} nucleus largely interprets it in the shell model framework while still indicating,
albeit theoretically, development of deformation/ shapes associated with the excited states of the nucleus. \\

The present work is directed at the spectroscopy of the $^{65}$Zn nucleus. It has been previously studied in varied experiments (such as described
in refs. \cite{Nil74,Ban94,Yu00,Muk01}) that observed different aspects of single particle and collective degrees of freedom in the excitation scheme 
of the nucleus. One of the earliest works was that by Nilsson and Sawa \cite{Nil74} wherein the lowest states of the nucleus were ascribed to shell model 
configurations in a truncated model space, while indicating the evidences of coupling of a valence (odd) neutron to the vibrations of the even-even ($^{64}$Zn) core. 
The subsequent work by the same group \cite{Nea78} also indicated the states corresponding to coupling between the neutron $g_{9/2}$ orbital and the core.
In conjunction, the latter study put forth a qualitative proposition of the states resulting from an asymmetric rotor coupled to single particle excitations; 
it was, however, identified that some of the states did not fit this interpretation.
Similarly, Banerjee {\it{et al.}} \cite{Ban94} identified the lowest positive parity states, below $\approx$ 5 MeV, as originating 
from the coupling of an odd neutron in the $g_{9/2}$ orbital to the quadrupole excitations of the $^{64}$Zn (anharmonic vibrator) core.
The later work by Yu {\it{et al.}} \cite{Yu00}, on the other hand, was directed at the high-spin states of the nucleus and 
used state-of-the-art experimental facilities to report superdeformed and highly deformed bands therein. 
However, no transitions could be confirmed linking these bands to the low-lying states. 
Incidentally, the most \textquotedblleft recent\textquotedblright \cite{Muk01} nuclear structure study on $^{65}$Zn was more than two-decades ago 
that, primarily, established the $g_{9/2}$ rotational band and proton alignment therein. Along with, it identified a shape transition in the 
nucleus from being near-oblate at lower excitations to triaxial prolate at intermediate spins.  
The study, additionally, identified several new transitions and levels while extending the level scheme of the nucleus to excitation $\approx$ 10.5 MeV. 
The current investigation was undertaken in the light of recent \cite{Zha25,Aya22} and intriguing findings in the Zn isotopes. In particular, the 
identification of octupole correlation in $^{67}$Zn \cite{Zha25} provided an impetus to probe the associated structure in the neighboring nuclei. 
In the process, several changes have resulted in the existing excitation scheme of $^{65}$Zn and new features have been observed therein. The findings are elucidated 
through the subsequent sections of this paper. \\

\section{Experimental Details and Data Analysis}

\begin{figure}
\includegraphics[angle=0,scale=.35,trim=0.0cm 0.0cm 0.0cm 0.0cm,clip=true]{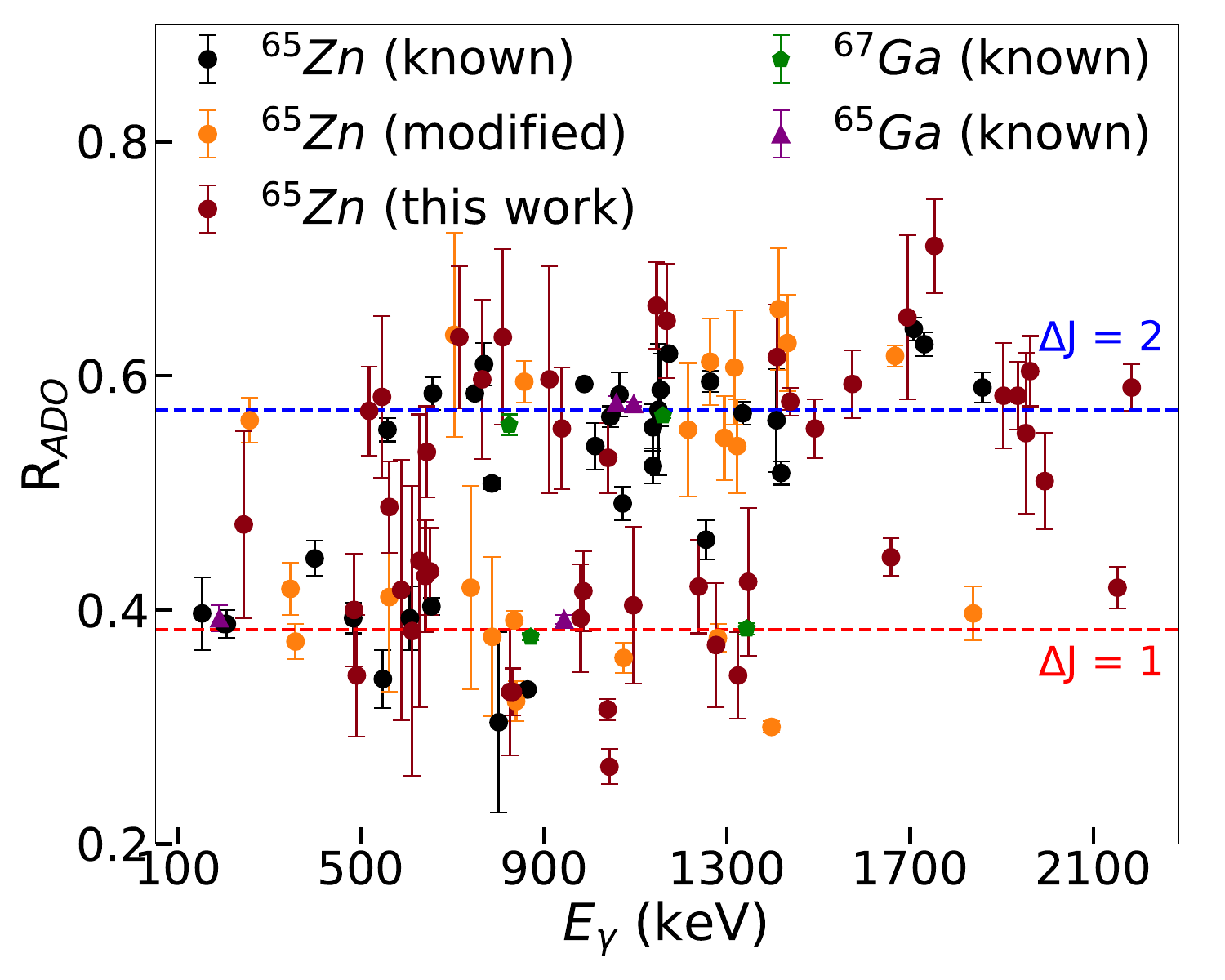}
\caption{\label{fig1}The $R_{ADO}$ values of transitions of $^{65}$Zn and those of other nuclei, used as reference, following the present analysis. 
The points corresponding to the \textquotedblleft known\textquotedblright transitions of $^{65}$Zn are the ones reported in the previous studies \cite{Muk01} and 
whose multipolarities as well as spin-parities of the associated levels remain unchanged in the present work. The transitions marked \textquotedblleft modified\textquotedblright
are the ones for which the multipolarities and/ or spin-parities of the initial and/ or the final level have been modified from the previous assignment. The transitions
indicated as \textquotedblleft this work\textquotedblright are newly observed in the present work and have been assigned the multipolarities herein.}
\end{figure}

\begin{figure}
\includegraphics[angle=0,scale=.28,trim=0.0cm 0.0cm 0.0cm 0.0cm,clip=true]{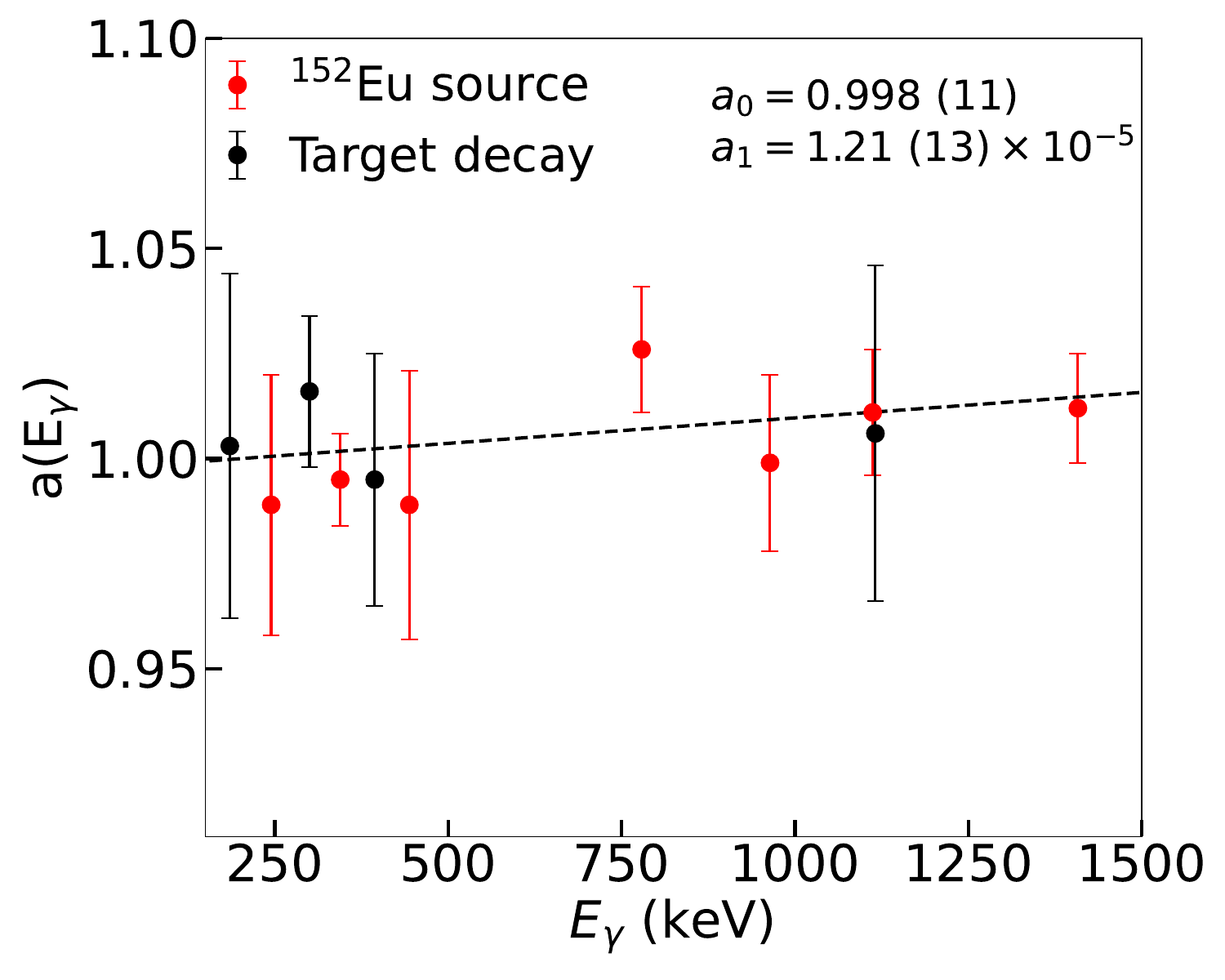}
\includegraphics[angle=0,scale=.28,trim=0.0cm 0.0cm 0.0cm 0.0cm,clip=true]{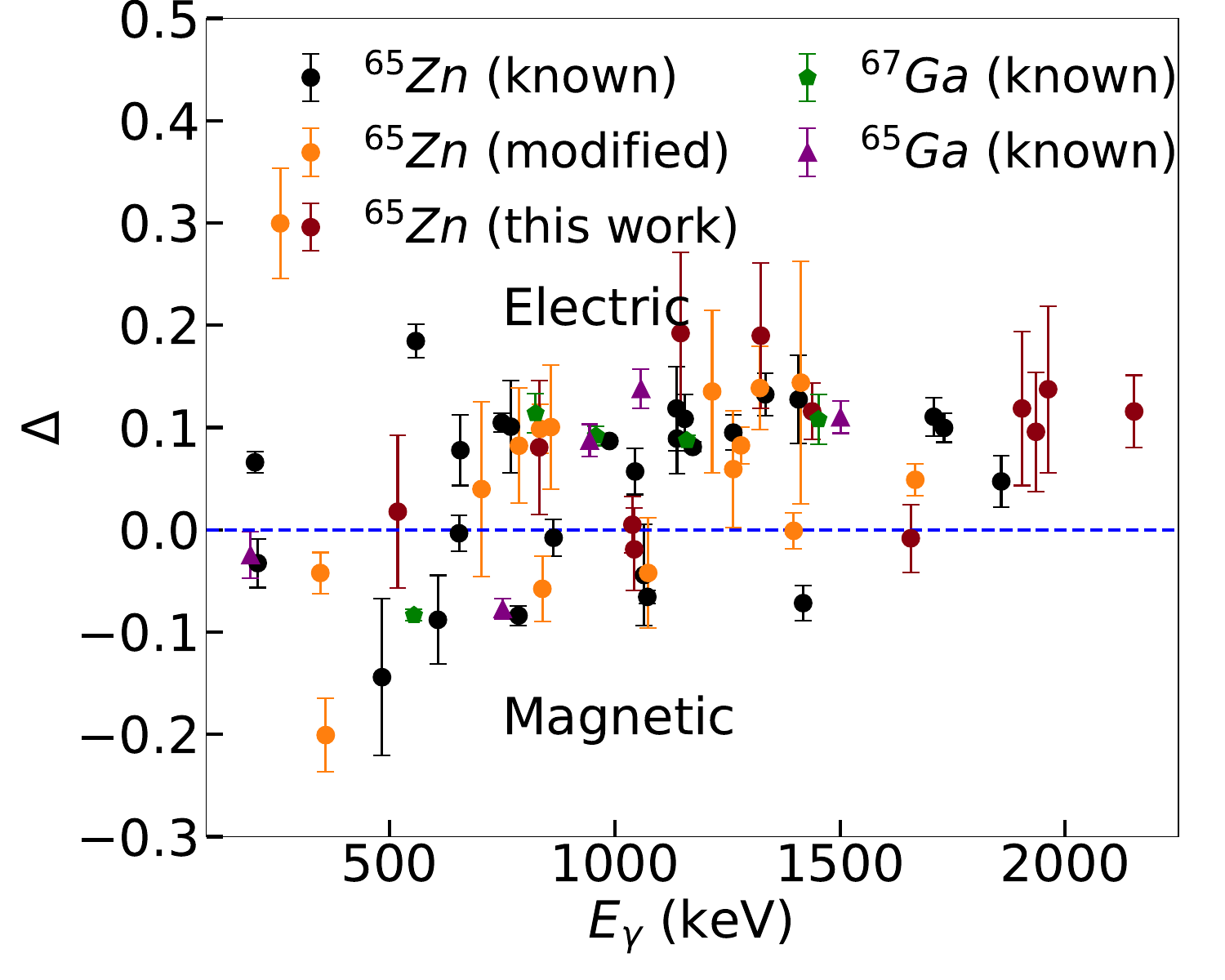}
\includegraphics[angle=0,scale=.28,trim=0.0cm 0.0cm 0.0cm 0.0cm,clip=true]{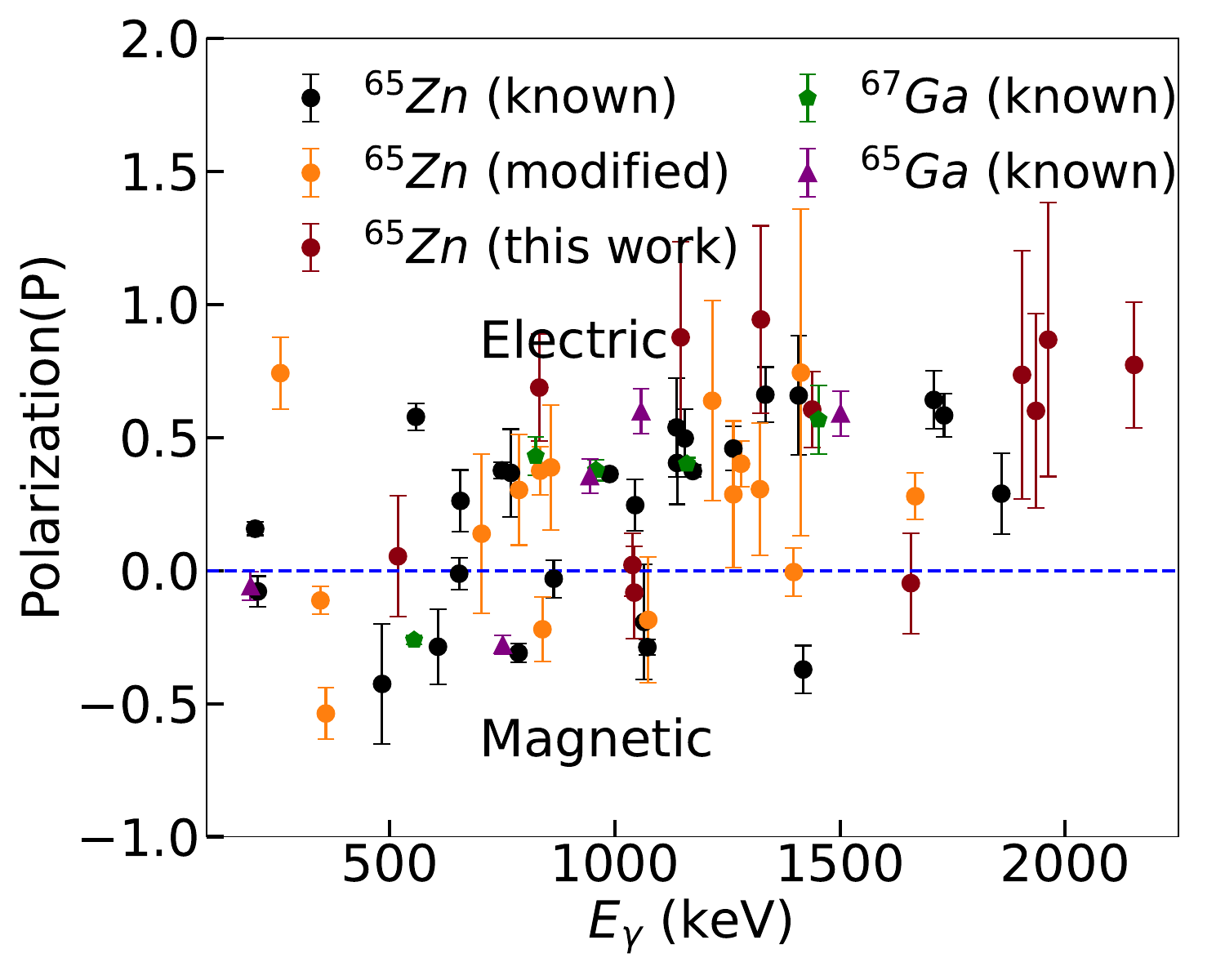}
\caption{\label{fig2}(a) Geometrical asymmetry ($a$) of the present setup as a function of $\gamma$-ray energies 
as obtained from the $^{152}$Eu source and from the offline decays of nuclei populated in the irradiation of the natural Cu target.
The plot includes the linear fit of the data and the parameters ($a_0, a_1$) therefrom.
(b) Polarization asymmetry ($\Delta$) of transitions of $^{65}$Zn, following the present work, along with those of the 
previously known transitions of other nuclei populated in the same experiment. (c) Linear 
polarization ($P$) of transitions of $^{65}$Zn along with those of other nuclei populated in the same experiment. 
The $\Delta$ and $P$ values for selected transitions of other isotopes, that were 
produced in the same experiment, are included for validation. The legends \textquotedblleft known\textquotedblright, \textquotedblleft modified\textquotedblright and 
\textquotedblleft this work\textquotedblright may be read as per the explanation in the caption of Fig.~1.}
\end{figure}

Excited states of the nucleus of interest were populated using $^{63}$Cu($\alpha$,pn)$^{65}$Zn reaction at E$_{lab}$ = 30 MeV. 
The target was 8.9 mg/cm$^2$ thick self-supporting foil of natural Cu ($^{63}$Cu $\approx$ 69\% and $^{65}$Cu $\approx$ 31\%).
The reaction with the $^{65}$Cu isotope did not contribute to the population of $^{65}$Zn owing to the threshold ($\approx$ 32 MeV) of the respective ($p3n$) channel. 
The $\alpha$-beam was delivered by the K130 Room Temperature Cyclotron (RTC) at the Variable Energy Cyclotron Centre (VECC) in Kolkata, India.
The Digital INGA (Indian National Gamma Array) setup at VECC was used as the detection system for the $\gamma$-rays emitted by the 
de-exciting nuclei. During the experiment the array consisted of 9 Compton suppressed HPGe clover detectors positioned at 
40$^\circ$ (2 detectors), 90$^\circ$ (5 detectors) and 125$^\circ$ (2 detectors). The pulse processing and data acquisition system 
was a digital one based on 12-bit, 250 MHz PIXIE-16 digitizer modules manufactured by XIA LLC, USA and running on a firmware 
conceptualized at the UGC-DAE CSR, Kolkata Centre \cite{Das18}. In-beam data, in time stamped listmode format, were acquired subject to an event trigger corresponding
to $\gamma$-$\gamma$ coincidence multiplicity, M$_\gamma$ $\ge$ 2. The two- and higher-fold coincidence events collected 
during the experiment were $\approx$ 1.7$\times$10$^9$. As is the practice, data, with event trigger set on singles (M$_\gamma$ $\ge$ 1) 
were also acquired with standard radioactive sources ($^{152}$Eu, $^{133}$Ba)  
for energy and efficiency calibration of the detection system as well as to determine its intrinsic geometrical asymmetry for polarization measurements (explained hereafter). \\

The coincidence data were sorted into symmetric and asymmetric (angle dependent) matrices using the IUCPIX \cite{Das18}
package developed at the UGC-DAE CSR, Kolkata Centre. A $\gamma$-$\gamma$-$\gamma$ cube was also constructed from the three- and higher-fold 
events, using the PIXSORT program of the IUCPIX package and the relevant subroutines 
of the INGASORT \cite{Bho01} code. The sorted data was analyzed using the RADWARE \cite{Rad95} and the CUBIX \cite{Dud25} codes 
towards determining the measurables relevant to the excitation scheme of the nucleus of interest. \\

The assignment of multipolarities of the $\gamma$-ray transitions followed their $R_{ADO}$ (Ratio of Angular Distribution from Oriented nuclei) values
using,

\begin{equation}
R_{ADO} = \frac{I_{\gamma1} \ at \ 125^\circ \ (Gated \ by \ \gamma_2 \ at \ all \ angles)}{I_{\gamma1} \ at \ 90^\circ \ (Gated \ by \ \gamma_2 \ at \ all \ angles)}
\end{equation}

\noindent{$I$ represents the intensity of the $\gamma_1$ transition, at the respective angle, as extracted from the spectrum corresponding to 
a gate on the $\gamma_2$ transition, detected at any angle. The asymmetric angle dependent matrices constructed for the purpose had $\gamma$-rays detected at any angle 
on the X-axis and the coincident ones, detected at 125$^\circ$ (90$^\circ$) on the Y-axis. As far as the reference values for the $R_{ADO}$ are concerned, 
those corresponded to 0.38$\pm$0.01 for stretched dipole ($\Delta$J = 1) transitions and 0.57$\pm$0.01 for the 
stretched quadrupole ($\Delta$J = 2) ones. These were concluded from the weighted average of the $R_{ADO}$ values corresponding to the previously known transitions 
of other nuclei, such as the $^{65,67}$Ga isotopes, that were considerably populated in the reaction used herein.  
Fig.~1 illustrates the $R_{ADO}$ values of different $\gamma$-ray transitions, observed in this study.} \\

The HPGe clover detector, as used in this work, is merited with the provision \cite{Duc99} for determining the electric/ magnetic nature of the $\gamma$-ray transitions
based on their linear polarization that bears on the probability of scattering of the $\gamma$-rays perpendicular or parallel to the reaction plane.
Each of the four crystals of clover detector operates as the scatterer and the adjacent ones as absorbers. 
The polarization asymmetry of a $\gamma$-ray transition is quantified by,

\begin{equation}
\Delta = \frac{aN_\perp \ - \ N_\parallel}{aN_\perp \ + \ N_\parallel}
\end{equation}

\noindent{where $N_\perp$ ($N_\parallel$) are the number of photons of the $\gamma$-ray
that are scattered perpendicular (parallel) to the reference plane that is defined by 
the direction of the beam for the reaction and the direction of emission of the $\gamma$-ray. 
Transitions are interpreted to be of electric (magnetic) nature following the positive (negative) values of $\Delta$; values 
near 0 are indicative of mixed electric/ magnetic character. 
The $N_\perp$ ($N_\parallel$) value of the $\gamma$-rays is extracted from asymmetric
matrix constructed with only the events corresponding to perpendicular (parallel) scattering in the detectors 
at 90$^o$ on the Y-axis and the coincident $\gamma$-rays detected at any angle on the X-axis. 
The coincidence conditions on the scattering events are directed at ensuring the unambiguous identification of the $\gamma$-rays of interest. 
The $a$ in Eq.~2 denotes the inherent geometrical asymmetry of the detector array, and is determined from the difference between $N_\perp$ and $N_\parallel$ 
for the $\gamma$-rays of unpolarized radioactive sources, such as $^{152}$Eu and the irradiated (natural Cu) target, and using $a = N_\parallel/N_\perp$.
The typical plot of $a$ as a function of $\gamma$-ray energies, as extracted for the present setup, is 
illustrated in Fig.~2(a). The data for $a$ is fitted with a first-order polynomial ($a_0~+~a_1x$) and its value is essentially 
the dominant offset ($a$ = $a_0$ = $a$ = 0.998$\pm$0.011 here) term; the slope ($a_1$) is typically insignificant (10$^{-5}$ in this case) and can be ignored for all practical purposes. 
Fig.~2(b) illustrates the polarization asymmetry ($\Delta$) of the transitions of $^{65}$Zn, observed in this study, along with the select ones of 
other nuclei ($^{65,67}$Ga), populated in the same experiment and used here to validate the analysis.   
It is noteworthy that the value of $\Delta$ fundamentally depends on the probability of scattering and is thus impacted by the energy of the transition. 
The dependence could be compensated by translating the polarization asymmetry ($\Delta$) into linear polarization $P$,

\begin{equation}
P = \frac{\Delta}{Q}
\end{equation}

\noindent{while using energy dependent polarization sensitivity $Q$, where}

\begin{equation}
Q(E_\gamma) = Q_0(E_\gamma)(CE_\gamma \ + \ D)
\end{equation}

\noindent{and,}

\begin{equation}
Q_0(E_\gamma) = \frac{\alpha + 1}{\alpha^2 + \alpha + 1}
\end{equation}

\noindent{$\alpha$ = $E_\gamma/m_ec^2$, with $m_ec^2$ being the electron rest mass energy, and $C$ and $D$ 
parameters are those reported by Palit {\it{et al.}} \cite{Pal00} as, $C = 0.000099 \ keV^{-1}$ and $D = 0.446$. 
Similar to the $\Delta$, the electric (magnetic) transitions are characterized by positive (negative) values of $P$; the mixed ones have $P$ values near zero.
Fig. 2(c) illustrates the $P$ values for the transitions of $^{65}$Zn and of $^{65,67}$Ga. The latter ones, as in case of $\Delta$, are for reference.} \\

The conventional methodologies for the experimental and the data analysis exercises for nuclear structure investigations 
have been used to validate, modify, and extend the levels scheme of $^{65}$Zn. The results are elucidated in the next section. 

\begin{figure*}
\includegraphics[angle=0,scale=0.77,trim=2.0cm 6.0cm 0.0cm 6.0cm,clip=true]{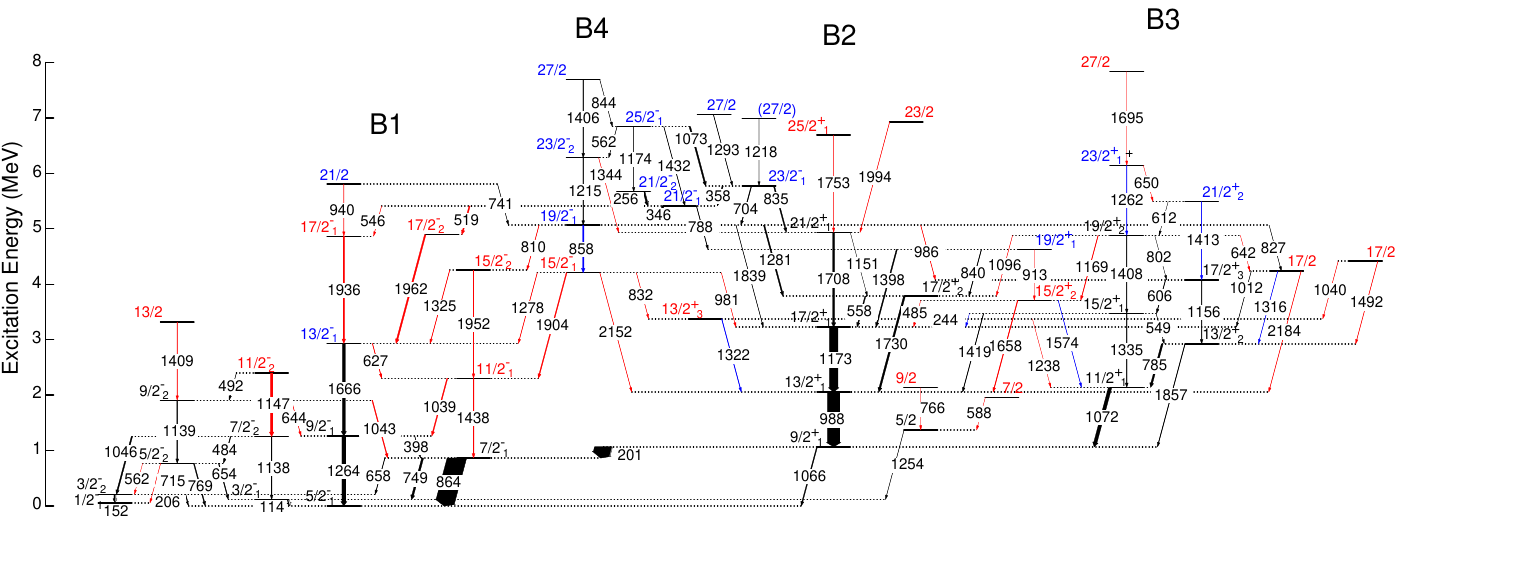}
\caption{\label{fig3} Excitation scheme of $^{65}$Zn following the present work. The $\gamma$-ray
transitions that have been newly identified in this study as well as the spin-parity labels of the new levels are indicated in red. The 
spin-parities that have been modified or confirmed vis-a-vis different or tentative assignment in the previous work \cite{Muk01} are indicated in blue. 
Similarly, the transitions for which the placements have been modified from that in the previous studies \cite{Muk01} are also indicated in blue.}
\end{figure*}

\section{Results}

\begin{figure}
\includegraphics[angle=0,scale=.27,trim=0.0cm 0.0cm 0.0cm 0.0cm,clip=true]{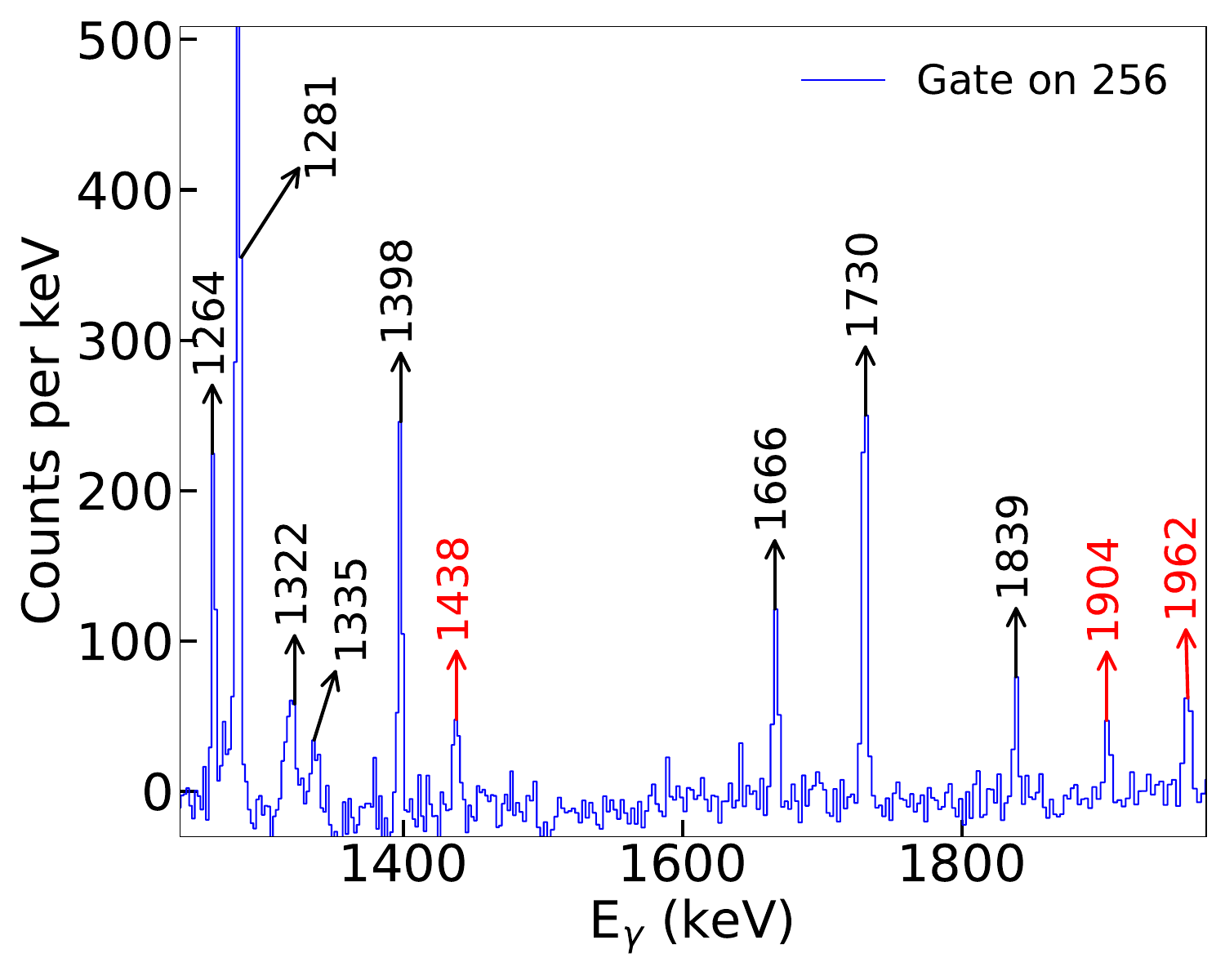}
\includegraphics[angle=0,scale=.27,trim=0.0cm 0.0cm 0.0cm 0.0cm,clip=true]{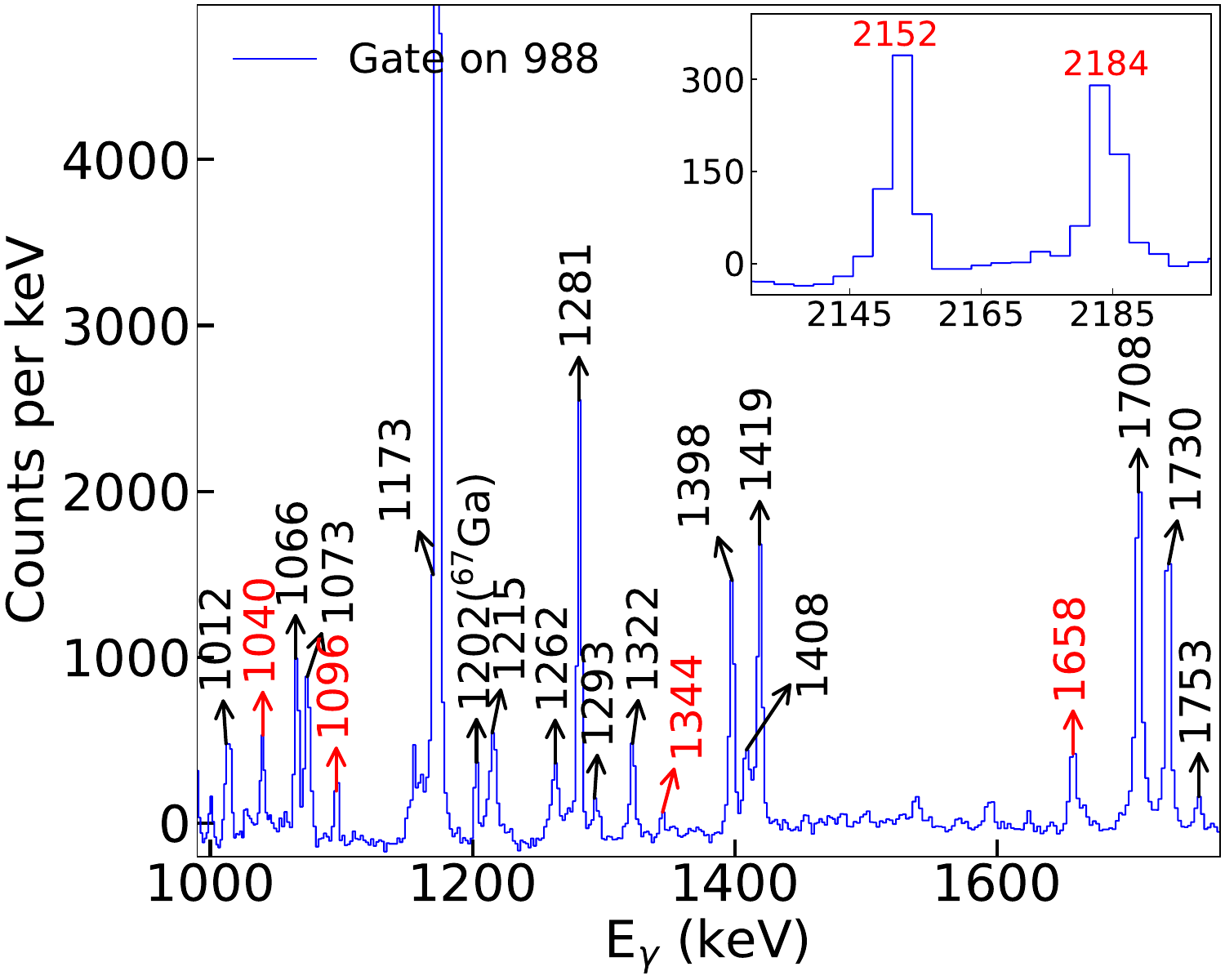}
\includegraphics[angle=0,scale=.27,trim=0.0cm 0.0cm 0.0cm 0.0cm,clip=true]{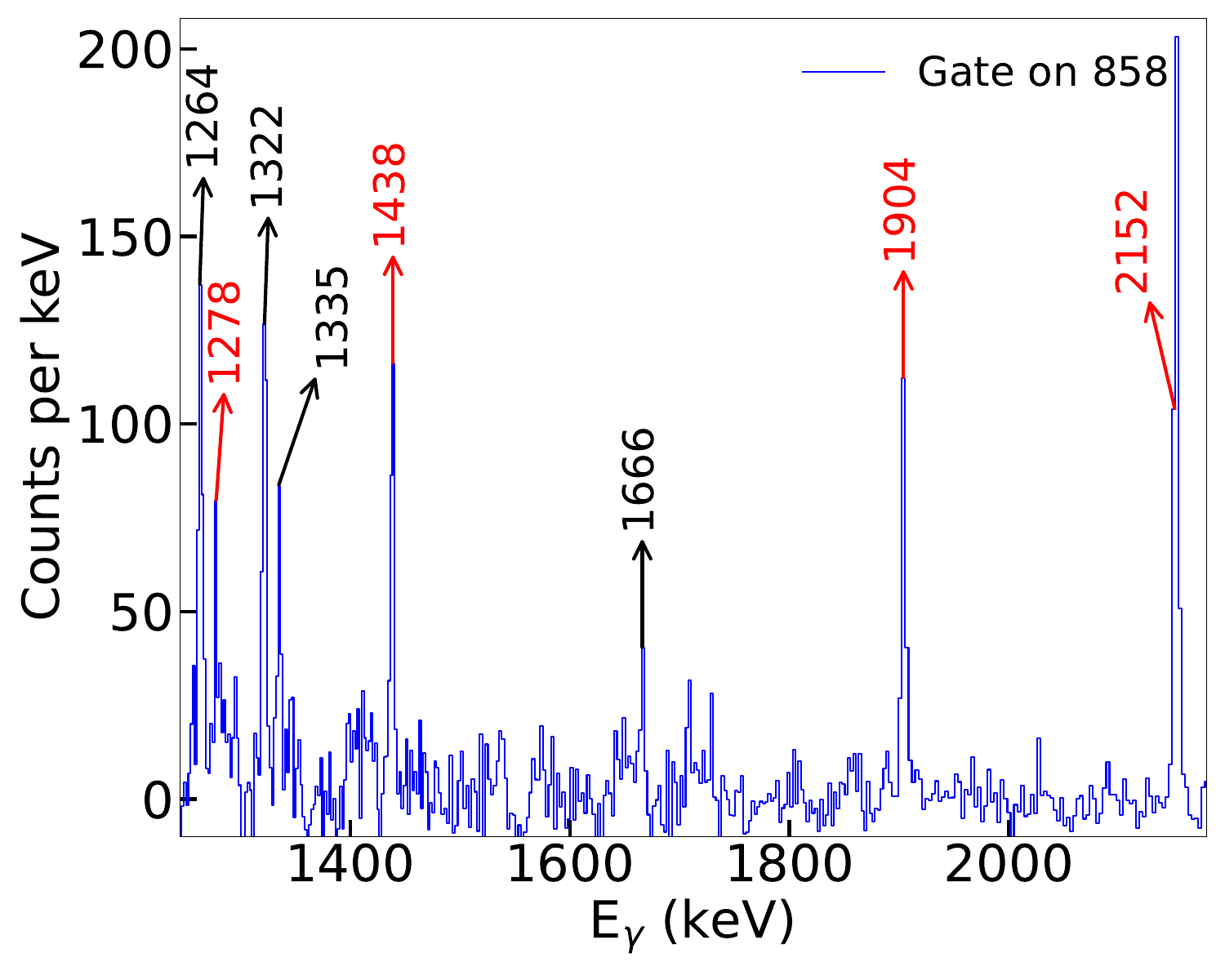}
\caption{\label{fig4}Representative spectra corresponding to gates applied on transitions of $^{65}$Zn, projected out of $\gamma$-$\gamma$-$\gamma$ cube. The respective gating transitions 
are indicated in the inset of the spectrum. The transitions newly identified in this work are marked in red.}
\end{figure}

\begin{figure}
\includegraphics[angle=0,scale=.27,trim=0.0cm 0.0cm 0.0cm 0.0cm,clip=true]{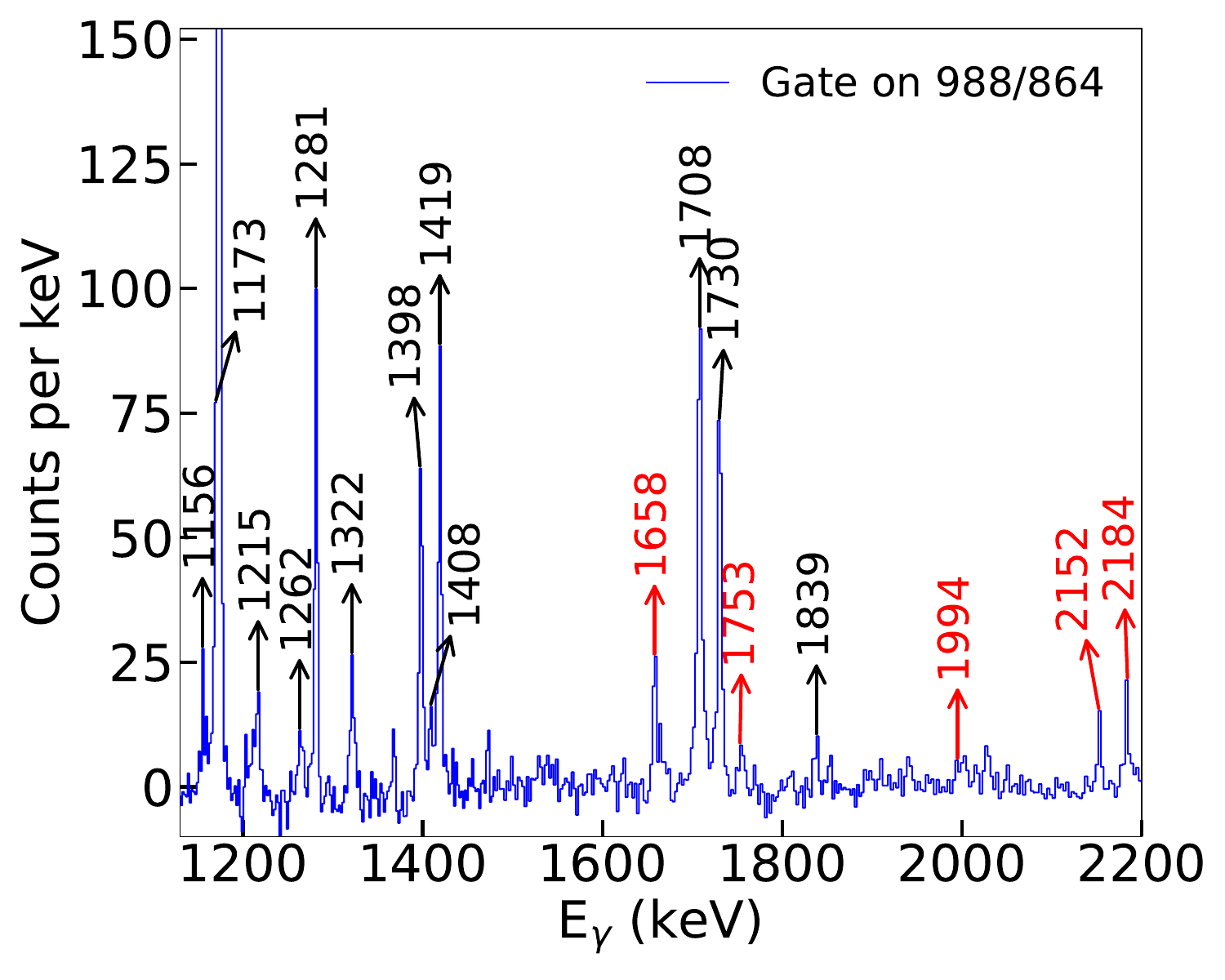}
\includegraphics[angle=0,scale=.27,trim=0.0cm 0.0cm 0.0cm 0.0cm,clip=true]{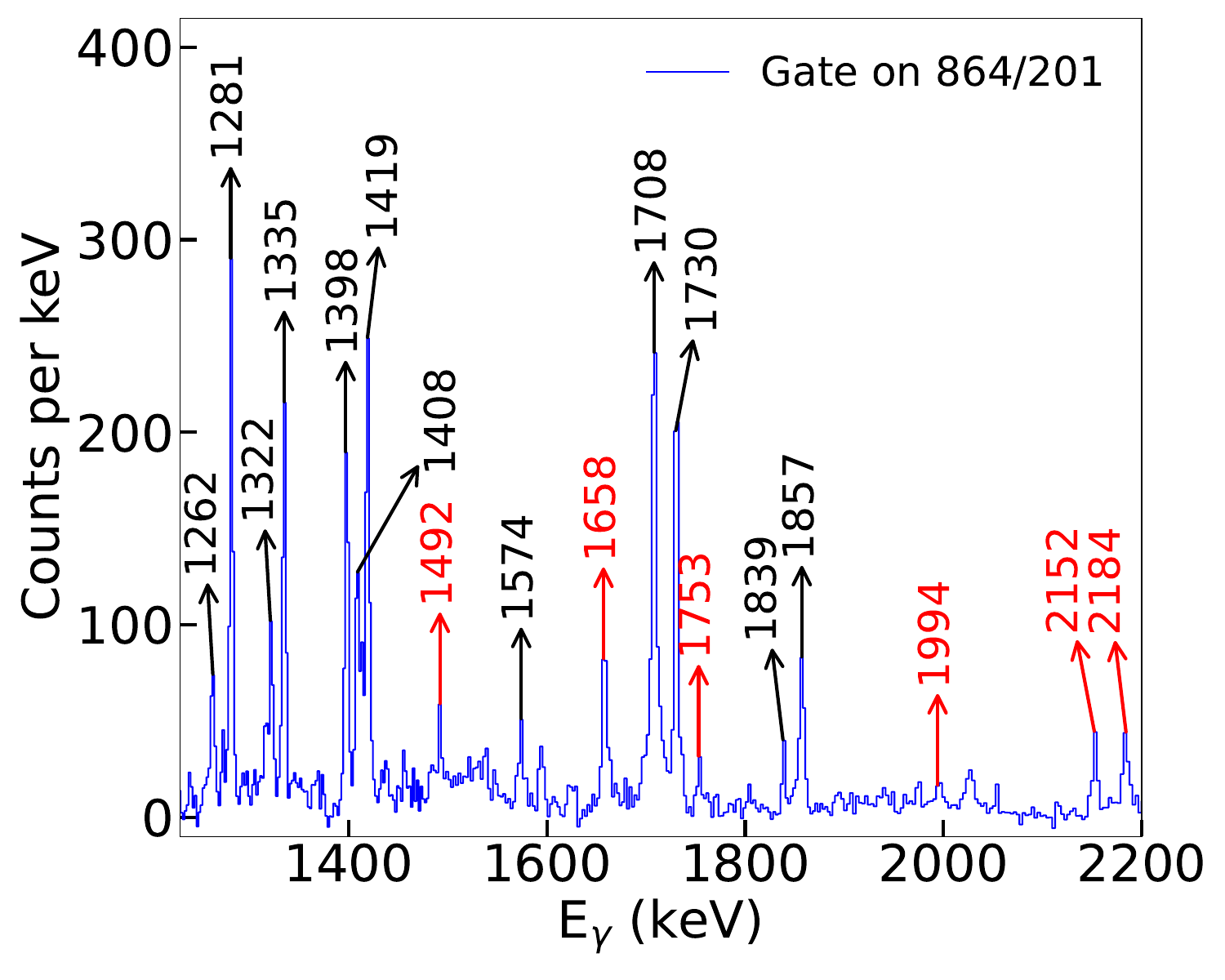}
\caption{\label{fig5}Representative spectra projected out of $\gamma$-$\gamma$-$\gamma$ cube corresponding to double gates set on transitions of $^{65}$Zn. 
These facilitate more stringent validation of the $\gamma$-$\gamma$ coincidences identified for the nucleus. The new transitions are indicated in red.}
\end{figure}

The excitation scheme of the $^{65}$Zn nucleus, as obtained in the present endeavor, is illustrated in Fig.~3. 
It follows the $\gamma$-$\gamma$ coincidence relationships along with the multipolarity of the transitions derived in this analysis. 
Figs. ~ 4 and 5 are representative gated spectra that facilitated the identification and/ or validation of the 
$\gamma$-$\gamma$ coincidences that constitute the level scheme.  
Table~1 lists the energy levels and $\gamma$-ray transitions of the $^{65}$Zn nucleus that have been newly identified  
and/ or have been confirmed in this work; the table also lists the properties (intensities, $R_{ADO}$, Polarization, multipolarity assignments) of the transitions,
following this analysis. 
This study has established 40 new $\gamma$-ray transitions in the level scheme of $^{65}$Zn while the placements of 7 (244-, 858-, 1262-, 1316-, 1322-, 1413- and 1574-keV) transitions 
have been modified from those in the previous study \cite{Muk01}. 
There are 11 (440-, 510-, 935-, 1163-, 1227-, 1349-, 1459-. 1476-, 1608-, 1656-, and 1916-keV) transitions, 
previously reported \cite{Muk01} for the nucleus, that could not be confirmed in this investigation.  
The level scheme of the nucleus has been established up to an excitation energy E$_x$ $\approx$ 8 MeV and spin $\approx$ 14$\hbar$.
The spin-parities of several levels, particularly those 
above $\approx$ 5 MeV, have either been modified from the existing assignments or have been newly assigned following the 
analysis herein.
It is noteworthy that the immediately previous \cite{Muk01} spectroscopic study on the nucleus, albeit around three-decades ago, did not include 
polarization measurement of the $\gamma$-ray transitions. The spin-parities of the levels thus followed 
\textquotedblleft DCO-type analysis\textquotedblright and were based on the known value of the spin-parity for the ground state. 
The setup for the present experiment, further to provisions for measurement of angular distribution/ correlation of the $\gamma$-ray transitions, 
also facilitated determination of their electric/ magnetic nature based on the linear polarization. 
These measurements translated into more definite multipolarity assignments of the transitions and spin-parity assignments of the levels in this study.   \\ 

This analysis has identified/ validated multiple band structures in the excitation scheme of $^{65}$Zn. 
These have been labeled as B1-B4 in Fig.~3. 
The ground state 5/2$^-$ band (B1)
consists of 1264-, 1666-, 1936- and 940-keV transitions of which the latter two have been newly identified in the present work.
In fact, even the 1264- and the 1666-keV transitions were not reported in the last 
study on this nucleus; the work was principally directed at its high-spin states and positive parity band structures. 
These transitions, however, were identified in the earlier works \cite{Nil74,Nea78} and have also been confirmed herein.  
The 5/2$^-$ band in $^{65}$Zn is similar 
to that in the neutron richer odd-A isotope $^{67}$Zn \cite{Zha25}; the intra-band transitions, in the two, have energies of the same order of magnitude. The 
band in $^{65}$Zn, however, is extended by one state in the present work, albeit with limited population, vis-a-vis the sequence in $^{67}$Zn. 
It is interesting to note that the ground state of the neutron deficient odd-A isotope $^{63}$Zn is 3/2$^-$, presumably based on the odd neutron occupying the
$p_{3/2}$ orbital. The change of the ground state spin-parity to 5/2$^-$ in $^{65}$Zn and heavier (odd-A) isotopes could indicate their evolving 
(ground state) deformation that drives the down slope of the $f_{5/2}$ orbital and favors its occupancy over the $p_{3/2}$. These observations 
provide the impetus to probe the deformation characteristics associated with the band structures across Zn isotopes and the same has been 
addressed in the next section. \\  

The 9/2$^+$ positive parity band (B2) of $^{65}$Zn has been considerably modified in the present study. 
It is the principally intense structure of the excitation scheme that de-excites to the ground state by 201- and 864-keV transitions.
The transitions 988-, 1173- and 1708-keV of the band were 
previously \cite{Muk01} known and have been confirmed herein. The next higher transition 1753-keV has been newly identified in this study. This work could not confirm 
any higher states for the band. The transitions 
835- and 1073-keV that were previously placed by Mukherjee {\it{et al.}} \cite{Muk01} as members of the 9/2$^+$ band have now been assigned different multipolarities, following the current 
analysis, and thus excluded from the sequence. The ADO and the P (polarization) values for the 835-keV $\gamma$-ray indicate it to be an E1 transition; 
Fig.~6 illustrates the difference in the 835-keV transition energy peaks corresponding to the perpendicular and the parallel scattering events 
in the 90$^\circ$ detectors that translates into the positive value of its polarization (as per Eqs.~2 and 3) and vindicates its electric nature. 
Similarly, the ADO and the polarization values of the 1073-keV transition uphold its M1 nature, at variance with the previous assignment 
and thus the transition is not a member of the 9/2$^+$ band. Mukherjee {\it {et al.}} had also reported 1155-, 1227-, and 1349-keV transitions 
in the sequence. The 1155-keV (1156-keV, here) transition is also present in the 11/2$^+$ band, as reported by Mukherjee {\it {et al.}} and as confirmed 
in the present work. There is no evidence of any additional 1155-keV $\gamma$-ray has in the excitation scheme of the $^{65}$Zn nucleus. 
There is a sharp drop in the intensity of the intra-band transitions across the 17/2$^+$ state. This is similar to the observations reported by 
Mukherjee {\it{et al.}} \cite{Muk01}. The intensity above the 17/2$^+$ level branches out into several transitions that mostly overlap 
between the present work and the previous studies \cite{Muk01}, except for three differences. First, the placement of the 1322-keV transition has been modified 
from that assigned by Mukherjee {\it{et al.}} \cite{Muk01}. Second, the 1476-keV transition that was identified as a new transition by 
Mukherjee {\it{et al.}} could not be validated in this analysis. And, finally, the transitions 981- and 484-keV have been newly identified in the present work. 
Further, the 
1227- and the 1349-keV transitions that were identified by Mukherjee {\it {et al.}} as the two highest members of the 9/2$^+$ sequence, were also not observed 
in the level structure of $^{65}$Zn following this study. \\

The 11/2$^+$ band (B3) also has been modified from the previously \cite{Muk01} assigned structure. The placements of the 1262- and the 1413-keV transitions 
have been changed, along with modification of the spin-parities of the respective initial states. The band has been extended by the newly observed 1695-keV transition,
though without any parity assignment of the corresponding new level at 7.8 MeV. The latter is the highest state of the level scheme, observed in this investigation. \\

The negative parity band (B4) based on the 15/2$^-$ state bears a \textquotedblleft close\textquotedblright resemblance with that observed in the neighboring $^{67}$Zn \cite{Zha25}
isotope and identified as evidence of octupole correlations therein. 
As has been pointed out in Ref.~\cite{Zha25} the evolution of the excitation energy of the 15/2$^-$ state, relative to that of the 9/2$^+$, across the odd-A Zn isotopes
should mimic that of the 3$^-$ state in the even-A (Zn) core. The energy of the 15/2$^-$ state that has been newly identified in the present work follows 
this trend as illustrated in Fig.~7 (similar to Fig.~3(a) in Ref.~\cite{Zha25}). 
Additionally, the excitation energies of the 19/2$^-$, the 23/2$^-$, and the 27/2$^-$ states of the band (B4) in $^{65}$Zn also align with the same trend vis-a-vis 
their energies in the $^{67}$Zn isotope. 
The present sequence in $^{65}$Zn is built on the modified placement of the 
858-keV transition \cite{Muk01}. The spin-parities of the initial and the final levels of the transition have been 
fixed on the basis of its ADO and polarization (P) values that indicate the 858-keV transition to be of E2 multipolarity; the previous work by Mukherjee {\it{et al.}} 
\cite{Muk01} reported the same assignment for the transition. 
The other two transitions of the band, 1215- and 1406-keV were known \cite{Muk01} ones. However, the spin-parities of the respective (de-excited) initial 
levels have been modified in this work. The ADO and the P values of the 1215-keV $\gamma$-ray conform with those of an E2 multipolarity and that overlaps with the 
previous assignment by Mukherjee {\it{et al.}} \cite{Muk01}. 
The 1406-keV transition, 
however, is too weak in this data for determining the polarization; it has, nevertheless, been identified as a quadrupole (Q) transition based on the ADO value 
and the spin of the respective level has been assigned accordingly.  
The 15/2$^-$ level at the band head of the B4 sequence
de-excites by 2152- and 981-keV transitions to the 9/2$^+$ band. These have been newly identified in the present investigation. The 2152-keV transition has been assigned 
E1 multipolarity following the respective ADO and P values. The intensity of the 981-keV transition is modest and consequently its polarization value could not be 
extracted; the ADO value implies that the transition is of dipole (D) nature. However, the spin-parities of its initial and final states could be confirmed 
from other, more intense, transitions and thus the 981-keV transition could be interpreted to be of E1 multipolarity. The 15/2$^-$ state also de-excites 
by the 1904-keV $\gamma$-ray that has been newly identified in this study and has been assigned an E2 multipolarity based on its ADO and P values.
Additionally, the 15/2$^-$ level also de-excites by the 1278-keV transition to the ground state band (B1) and by the 832-keV transition to the non-yrast 13/2$^+$ state. 
The 1278-keV transition is a new one, following this work. The ADO value indicates it to be a dipole (D) but the transition is weakly intense for any polarization (P) analysis.
The 832-keV transition too is a weak transition that has been first identified in the present analysis.
Further, there are transitions branching out of the B4 sequence from its higher states as well. The 810- and the 1344-keV transitions are the newly identified ones that
respectively de-excite the 19/2$^-$ state and the 23/2$^-$ state of the band. These $\gamma$-rays are weakly intense for polarization analysis but the respective ADO ratios 
indicate the 810-keV transition to be a quadrupole (Q) and the 1344-keV transition to be a dipole (D). The top most state of the sequence that is tentatively assigned spin $J$ = 27/2
also branch out by a weak, previously known \cite{Muk01} 844-keV transition. \\

In addition to the aforesaid bands, there are other non-yrast states and sequences that characterize the excitation scheme of the $^{65}$Zn nucleus emerging from the present study.
The use of $\alpha$-beam is known to facilitate the population of such structures.
Some of these and the associated $\gamma$-ray transitions were previously known \cite{Muk01} while others have been newly identified in this work.  
The next section addresses the excitation mechanisms and the associated parameters underlying the bands and other structures in the excitation scheme. \\

\LTcapwidth=\textwidth
\begin{longtable*}{ccccccccccc}
\caption{\label{tab1}Details of the levels and the $\gamma$-ray transitions of $^{65}$Zn observed/ validated in the present study. The level energies ($E_i$ and $E_f$) are those following the least-square fit using the GTOL code under the Nuclear Data Services framework of the IAEA \cite{iaea}. The $D$ and $Q$ in the column for multipolarity assignments respectively represents dipole and quadrupole transitions for which the polarization measurement could not be carried out.} \\
\hline
$E_i (keV) $   & $E_{\gamma} (keV) $     & $E_f (keV) $   &$I_{\gamma}$ & $J_i^{\pi}$     & $J_f^{\pi}$&$R_{ADO}$  &$\Delta_{pol}$ & P & Multipolarity\\
\hline
\hline
\endfirsthead

\multicolumn{11}{c}%
{{ \tablename \thetable{} -- continued from previous page}} \\
\hline
$E_i (keV) $   & $E_{\gamma} (keV) $    & $E_f (keV) $   & $I_{\gamma}$ &  $J_i^{\pi}$ & $J_f^{\pi}$&$R_{ADO}$ &$\Delta_{pol}$ & P & Multipolarity\\
\hline
\endhead

\hline
\multicolumn{11}{c}{Continued in next page}\\
\hline
\endfoot
\endlastfoot

54.0(5)        & ~                     & 0               & ~            & 1/2$^{-}$    & 5/2$^{-}$  & ~         & ~            &  ~        & ~            \\
114.6(3)       & 114.1(5)              & 0               & 51(1)        & 3/2$^{-}$    & 5/2$^{-}$  & 0.41(1)   & ~            & ~         & D \\
206.5(3)       & 152.2(5)              & 54              & 72(3)        & 3/2$^{-}$    & 1/2$^{-}$  & 0.40(3)   & ~            & ~         & D \\
~              & 206.2(5)              & 0               & 28(1)        & 3/2$^{-}$    & 5/2$^{-}$  & 0.39(1)   & -0.03(2)     & -0.08(6)  & M1+E2 \\
768.6(3)       & 561.6(5)              & 206             & 2(1)         & 5/2$^{-}$    & 3/2$^{-}$  & 0.49(4)   & ~            & ~         & D+Q \\
~              & 653.5(5)              & 115             & 40(1)        & 5/2$^{-}$    & 3/2$^{-}$  & 0.40(1)   & -0.003(18)   & -0.012(60)& M1+E2 \\
~              & 714.8(5)              & 54              & 6(1)         & 5/2$^{-}$    & 1/2$^{-}$  & 0.63(6)   & ~            & ~         & Q \\
~              & 768.7(5)              & 0               & 52(2)        & 5/2$^{-}$    & 5/2$^{-}$  & 0.61(2)   &  0.10(5)     &  0.37(16) & $\Delta$J = 0 \\
864.3(3)       & 657.5(5)              & 206             & 24(1)        & 7/2$^{-}$    & 3/2$^{-}$  & 0.59(1)   &  0.08(3)     &  0.26(12) & E2 \\
~              & 749.2(5)              & 115             & 113(2)       & 7/2$^{-}$    & 3/2$^{-}$  & 0.59(1)   &  0.10(1)     &  0.38(3)  & E2 \\
~              & 864.4(5)              & 0               & 1125(23)     & 7/2$^{-}$    & 5/2$^{-}$  & 0.33(1)   & -0.01(2)     & -0.03(7)  & M1+E2 \\
1065.5(4)      & 200.6(5)              & 864             & 1000         & 9/2$^{+}$    & 7/2$^{-}$  & 0.39(1)   &  0.07(1)     &  0.16(2)  & E1 \\
~              & 1065.8(5)             & 0               & 54(1)        & 9/2$^{+}$    & 5/2$^{-}$  & 0.58(2)   & -0.04(5)     & -0.19(22) & M2+E3 \\
1252.5(4)      & 483.5(5)              & 769             & 34(1)        & 7/2$^{-}$    & 5/2$^{-}$  & 0.39(1)   & -0.14(8)     & -0.43(23) & M1 \\
~              & 1046.1(5)             & 206             & 74(3)        & 7/2$^{-}$    & 3/2$^{-}$  & 0.57(1)   &  0.06(2)     &  0.25(10) & E2 \\
~              & 1138.2(5)             & 115             & 30(1)        & 7/2$^{-}$    & 3/2$^{-}$  & 0.56(2)   &  0.12(4)     &  0.54(19) & E2 \\
1263.3(3)      & 398.5(5)              & 864             & 15(1)        & 9/2$^{-}$    & 7/2$^{-}$  & 0.44(2)   & ~            & ~         & D+Q \\
~              & 1263.6(5)             & 0               & 233(8)       & 9/2$^{-}$    & 5/2$^{-}$  & 0.60(1)   &  0.09(2)     &  0.46(8)  & E2 \\
1369.1(6)      & 1254.4(5)             & 115             & 8(1)         & 5/2          & 3/2$^{-}$  & 0.46(2)   & ~            & ~         & D+Q \\
1907.5(4)      & 643.9(5)              & 1263            & 12(1)        & 9/2$^{-}$    & 9/2$^{-}$  & 0.54(4)   & ~            & ~         & $\Delta$J = 0 \\
~              & 1043.4(5)             & 864             & 45(4)        & 9/2$^{-}$    & 7/2$^{-}$  & 0.27(2)   & -0.02(4)     & -0.08(17) & M1+E2 \\
~              & 1138.9(5)             & 769             & 46(4)        & 9/2$^{-}$    & 5/2$^{-}$  & 0.52(2)   &  0.09(3)     &  0.40(16) & E2 \\
1956.9(8)      & 587.8(5)              & 1369            & 2(1)         & 7/2          & 5/2        & 0.42(11)  & ~            & ~         & D+Q \\
2053.6(4)      & 988.1(5)              & 1066            & 740(11)      & 13/2$^{+}$   & 9/2$^{+}$  & 0.59(1)   &  0.09(1)     &  0.36(1)  & E2 \\
2134.9(8)      & 765.8(5)              & 1369            & 4(1)         & 9/2          & 5/2        & 0.60(7)   & ~            & ~         & Q \\
2137.3(5)      & 1072.1(5)             & 1066            & 198(5)       & 11/2$^{+}$   & 9/2$^{+}$  & 0.46(3)   & -0.04(3)     & -0.17(12) & M1+E2 \\
2302.7(4)      & 1039.5(5)             & 1263            & 55(2)        & 11/2$^{-}$   & 9/2$^{-}$  & 0.32(1)   &  0.01(3)     &  0.02(12) & M1+E2 \\
~              & 1438.4(5)             & 864             & 42(1)        & 11/2$^{-}$   & 7/2$^{-}$  & 0.58(1)   &  0.12(3)     &  0.61(14) & E2 \\
2399.3(5)      & 491.7(5)              & 1908            & 10(1)        & 11/2$^{-}$   & 9/2$^{-}$  & 0.34(5)   & ~            & ~         & D \\
~              & 1146.9(5)             & 1252            & 138(3)       & 11/2$^{-}$   & 7/2$^{-}$  & 0.66(4)   &  0.19(8)     &  0.88(36) & E2 \\
2922.7(5)      & 785.4(5)              & 2137            & 104(3)       & 13/2$^{+}$   & 11/2$^{+}$ & 0.51(1)   & -0.02(3)     & -0.07(10) & M1+E2 \\
~              & 1856.9(5)             & 1066            & 29(1)        & 13/2$^{+}$   & 9/2$^{+}$  & 0.59(1)   &  0.05(2)     &  0.29(15) & E2 \\
2929.7(4)      & 627.2(5)              & 2303            & 23(1)        & 13/2$^{-}$   & 11/2$^{-}$ & 0.44(13)  & ~            & ~         & D+Q \\
~              & 1666.4(5)             & 1263            & 182(7)       & 13/2$^{-}$   & 9/2$^{-}$  & 0.62(1)   &  0.05(2)     &  0.28(9)  & E2 \\
3226.5(5)      & 1172.9(5)             & 2054            & 509(10)      & 17/2$^{+}$   & 13/2$^{+}$ & 0.62(1)   &  0.08(1)     &  0.37(2)  & E2 \\
3316.7(6)      & 1409.2(5)             & 1908            & 22(1)        & 13/2         & 9/2$^{-}$  & 0.62(5)   & ~            & ~         & Q \\
3375.3(6)      & 1238.4(5)             & 2137            &  4(1)        & 13/2$^{+}$   & 11/2$^{+}$ & 0.42(4)   &              &           & D             \\
               & 1321.8(5)             & 2054            & 26(1)        & 13/2$^{+}$   & 13/2$^{+}$ & 0.54(4)   &  0.14(4)     &  0.69(20) & $\Delta$J = 0 \\
3471.9(5)      & 244.2(5)              & 3226            & 2(1)         & 15/2$^{+}$   & 17/2$^{+}$ & 0.47(8)   & ~            & ~         & D+Q \\
~              & 548.9(5)              & 2923            & 7(1)         & 15/2$^{+}$   & 13/2$^{+}$ & 0.34(3)   & ~            & ~         & D \\
~              & 1334.9(5)             & 2137            & 39(1)        & 15/2$^{+}$   & 11/2$^{+}$ & 0.57(1)   &  0.13(2)     &  0.66(10) & E2 \\
~              & 1418.9(5)             & 2054            & 28(1)        & 15/2$^{+}$   & 13/2$^{+}$ & 0.52(1)   & -0.03(2)     & -0.15(10)& M1+E2 \\
3711.1(5)      & 484.7(5)              & 3226            & 2(1)         & 15/2$^{+}$   & 17/2$^{+}$ & 0.40(5)   & ~            & ~        & D \\
~              & 1573.9(5)             & 2137            & 13(1)        & 15/2$^{+}$   & 11/2$^{+}$ & 0.59(3)   & ~            & ~        & Q \\
~              & 1657.6(5)             & 2054            & 35(4)        & 15/2$^{+}$   & 13/2$^{+}$ & 0.45(2)   & -0.01(3)     & -0.05(19)& M1+E2 \\
3783.9(5)      & 557.5(5)              & 3226            & 59(1)        & 17/2$^{+}$   & 17/2$^{+}$ & 0.55(1)   &  0.18(2)     &  0.58(5) & $\Delta$J = 0 \\
~              & 1730.3(5)             & 2054            & 106(2)       & 17/2$^{+}$   & 13/2$^{+}$ & 0.63(1)   &  0.10(1)     &  0.58(8) & E2 \\
4078.4(5)      & 606.0(5)              & 3472            & 26(1)        & 17/2$^{+}$   & 15/2$^{+}$ & 0.39(3)   & -0.09(4)     & -0.29(14)& M1 \\
~              & 1155.7(5)             & 2923            & 37(1)        & 17/2$^{+}$   & 13/2$^{+}$ & 0.59(3)   &  0.11(2)     &  0.50(11)& E2 \\
4206.8(4)      & 831.8(5)              & 3375            & 9(1)         & 15/2$^{-}$   & 13/2$^{+}$ & 0.33(2)   &  0.08(6)     &  0.31(25)& E1 \\
~              & 980.7(5)              & 3226            & 7(1)         & 15/2$^{-}$   & 17/2$^{+}$ & 0.39(5)   & ~            & ~        & D \\
~              & 1277.6(5)             & 2930            & 17(1)        & 15/2$^{-}$   & 13/2$^{-}$ & 0.37(5)   & ~            & ~        & D \\
~              & 1904.1(5)             & 2303            & 51(2)        & 15/2$^{-}$   & 11/2$^{-}$ & 0.58(5)   &  0.12(8)     &  0.74(47)& E2 \\
~              & 2152.1(5)             & 2054            & 12(1)        & 15/2$^{-}$   & 13/2$^{+}$ & 0.42(2)   &  0.12(4)     &  0.77(24)& E1 \\
4238.4(5)      & 1012.1(5)             & 3226            & 10(1)        & 17/2         & 17/2$^{+}$ & 0.54(2)   & ~            & ~        & $\Delta$J = 0 \\
~              & 1315.5(5)             & 2923            & 12(1)        & 17/2         & 13/2$^{+}$ & 0.61(5)   & ~            & ~        & Q \\
               & 2183.8(5)             & 2054            & 13(1)        & 17/2         & 13/2$^{+}$ & 0.59(2)   & ~            & ~        & Q \\ 
4254.4(5)      & 1324.7(5)             & 2930            & 20(1)        & 15/2$^{-}$   & 13/2$^{-}$ & 0.34(4)   & ~            & ~        & D \\
~              & 1951.5(5)             & 2303            & 22(1)        & 15/2$^{-}$   & 11/2$^{-}$ & 0.55(7)   &  0.12(5)     &  0.79(29)& E2 \\
4414.8(7)      & 1039.8(5)             & 3375            &  7(1)        & 17/2         & 13/2$^{+}$ & 0.53(3)   & ~            & ~        & Q \\
               & 1492.0(5)             & 2923            & 13(1)        & 17/2         & 13/2$^{+}$ & 0.56(3)   & ~            & ~        & Q \\
4623.8(5)      & 839.7(5)              & 3784            & 29(1)        & 19/2$^{+}$   & 17/2$^{+}$ & 0.32(2)   & -0.06(3)     & -0.22(12)& M1 \\
~              & 912.8(5)              & 3711            & 2(1)         & 19/2$^{+}$   & 15/2$^{+}$ & 0.60(10)  & ~            & ~        & Q \\
~              & 1397.5(5)             & 3227            & 65(1)        & 19/2$^{+}$   & 17/2$^{+}$ & 0.30(1)   &  0.01(2)     & -0.01(9) & M1+E2 \\
4865.3(4)      & 1935.6(5)             & 2930            & 81(4)        & 17/2$^{-}$   & 13/2$^{-}$ & 0.58(3)   &  0.10(6)     &  0.60(37)& E2 \\
4880.2(5)      & 641.9(5)              & 4238            & 4(1)         & 19/2$^{+}$   & 17/2       & 0.43(5)   & ~            & ~        & D+Q \\
~              & 801.9(5)              & 4078            & 2(1)         & 19/2$^{+}$   & 17/2$^{+}$ & 0.30(8)   & ~            & ~        & D+Q \\
~              & 1096.5(5)             & 3784            & 6(1)         & 19/2$^{+}$   & 17/2$^{+}$ & 0.40(7)   & ~            & ~        & D+Q \\
~              & 1169.1(5)             & 3711            & 26(1)        & 19/2$^{+}$   & 15/2$^{+}$ & 0.65(5)   & ~            & ~        & Q \\
~              & 1408.4(5)             & 3472            & 31(1)        & 19/2$^{+}$   & 15/2$^{+}$ & 0.56(4)   &  0.13(4)     &  0.66(22)& E2 \\
4892.2(6)      & 1962.2(5)             & 2930            & 86(4)        & 17/2$^{-}$   & 13/2$^{-}$ & 0.60(3)   &  0.14(8)     &  0.87(51)& E2 \\
4934.4(5)      & 1150.7(5)             & 3784            & 4(1)         & 21/2$^{+}$   & 17/2$^{+}$ & 0.57(6)   &            ~ & ~        & Q \\
~              & 1707.8(5)             & 3226            & 128(3)       & 21/2$^{+}$   & 17/2$^{+}$ & 0.64(1)   &  0.11(2)     &  0.64(11)& E2 \\
5065.0(5)      & 810.4(5)              & 4254            & 17(1)        & 19/2$^{-}$   & 15/2$^{-}$ & 0.63(8)   & ~            & ~        & Q \\
        ~      & 826.5(5)              & 4238            & 7(1)         & 19/2$^{-}$   & 17/2       & 0.33(5)   & ~            & ~        & D \\
        ~      & 858.2(5)              & 4207            & 80(4)        & 19/2$^{-}$   & 15/2$^{-}$ & 0.60(2)   &  0.10(6)     &  0.39(23)& E2 \\
        ~      & 986.5(5)              & 4078            & 14(1)        & 19/2$^{-}$   & 17/2$^{+}$ & 0.42(3)   & ~            & ~        & D \\
        ~      & 1281.2(5)             & 3784            & 73(2)        & 19/2$^{-}$   & 17/2$^{+}$ & 0.38(1)   &  0.08(2)     &  0.40(9) & E1 \\
        ~      & 1838.6(5)             & 3226            & 13(1)        & 19/2$^{-}$   & 17/2$^{+}$ & 0.40(2)   & ~            & ~        & D \\
5411.3(5)      & 346.0(5)              & 5065            & 114(4)       & 21/2$^{-}$   & 19/2$^{-}$ & 0.42(2)   & -0.04(2)     & -0.11(5) & M1 \\
~              & 518.8(5)              & 4892            & 68(4)        & 21/2$^{-}$   & 17/2$^{-}$ & 0.57(4)   &  0.02(7)     &  0.05(23)& E2(+M3) \\
        ~      & 546.2(5)              & 4865            & 7(1)         & 21/2$^{-}$   & 17/2$^{-}$ & 0.58(7)   & ~            & ~        & Q \\
        ~      & 787.9(5)              & 4624            & 23(1)        & 21/2$^{-}$   & 19/2$^{+}$ & 0.38(7)   &  0.08(6)     &  0.30(21)& E1 \\
5491.8(6)      & 611.6(6)              & 4880            & 2(1)         & 21/2$^{+}$   & 19/2$^{+}$ & 0.38(12)  & ~            & ~        & D+Q \\
        ~      & 1413.1(5)             & 4078            & 18(1)        & 21/2$^{+}$   & 17/2$^{+}$ & 0.66(5)   &  0.14(12)    &  0.74(61)& E2 \\
5667.9(6)      & 255.9(5)              & 5411            & 95(3)        & 21/2$^{-}$   & 21/2$^{-}$ & 0.56(2)   &  0.30(5)     &  0.74(13)& $\Delta$J = 0 \\
5769.4(5)      & 357.7(5)              & 5411            & 65(2)        & 23/2$^{-}$   & 21/2$^{-}$ & 0.37(2)   & -0.20(4)     & -0.54(10)& M1 \\
~              & 704.3(5)              & 5065            & 38(1)        & 23/2$^{-}$   & 19/2$^{-}$ & 0.64(9)   &  0.04(9)     &  0.14(30)& E2 \\
        ~      & 835.2(5)              & 4934            & 66(1)        & 23/2$^{-}$   & 21/2$^{+}$ & 0.39(1)   &  0.10(2)     &  0.38(9) & E1 \\
5805.6(6)      & 740.6(5)              & 5065            & 9(1)         & 21/2         & 19/2$^{-}$ & 0.42(9)   & ~            & ~        & D+Q \\
        ~      & 940.2(5)              & 4865            & 15(1)        & 21/2         & 17/2$^{-}$ & 0.56(5)   & ~            & ~        & Q \\
6142.4(6)      & 650.3(5)              & 5492            & 3(1)         & 23/2$^{+}$   & 21/2$^{+}$ & 0.43(4)   & ~            & ~        & D+Q \\
        ~      & 1262.5(5)             & 4880            & 17(1)        & 23/2$^{+}$   & 19/2$^{+}$ & 0.61(4)   &  0.06(6)     &  0.29(28)& E2 \\
6280.3(6)      & 1215.1(5)             & 5065            & 38(1)        & 23/2$^{-}$   & 19/2$^{-}$ & 0.55(6)   &  0.14(8)     &  0.64(38)& E2 \\
        ~      & 1344.3(5)             & 4934            & 6(1)         & 23/2$^{-}$   & 21/2$^{+}$ & 0.42(6)   & ~            & ~        & D \\
6687.8(7)      & 1753.4(5)             & 4934            & 10(1)        & 25/2$^{+}$   & 21/2$^{+}$ & 0.71(4)   & ~            & ~        & Q \\
6842.2(5)      & 561.7(5)              & 6280            & 10(1)        & 25/2$^{-}$   & 23/2$^{-}$ & 0.41(8)   & ~            & ~        & D+Q \\
        ~      & 1072.6(5)             & 5769            & 84(3)        & 25/2$^{-}$   & 23/2$^{-}$ & 0.36(1)   & -0.04(5)     & -0.19(24)& M1 \\
	~      & 1173.7(5)             & 5668            & 17(1)        & 25/2$^{-}$   & 21/2$^{-}$ & ~         & ~            & ~        & (Q) \\
        ~      & 1431.7(5)             & 5411            & 18(1)        & 25/2$^{-}$   & 21/2$^{-}$ & 0.63(4)   & ~            & ~        & Q \\
6928.5(7)      & 1994.1(5)             & 4934            & 11(1)        & 23/2         & 21/2$^{+}$ & 0.51(4)   & ~            & ~        & D+Q \\
6987.8(7)      & 1218.4(5)             & 5769            & 7(1)         & (27/2)       & 23/2$^{-}$ & ~         & ~            & ~        & (Q) \\
7062.6(7)      & 1293.2(5)             & 5769            & 18(1)        & 27/2         & 23/2$^{-}$ & 0.55(4)   & ~            & ~        &  Q  \\
7686.6(6)      & 844.5(5)              & 6842            & 6(1)         & 27/2         & 25/2$^{-}$ & ~         & ~            & ~        & (D) \\
~              & 1406.4(5)             & 6280            & 32(2)        & 27/2         & 23/2$^{+}$ & 0.59(7)   & ~            & ~        &  Q \\
7837.3(8)      & 1694.8(5)             & 6142            & 4(1)         & 27/2         & 23/2$^{+}$ & 0.65(7)   & ~            & ~        &  Q \\
\hline
\bigskip
\end{longtable*}

\begin{figure}
\includegraphics[angle=0,scale=.27,trim=0.0cm 0.0cm 0.0cm 0.0cm,clip=true]{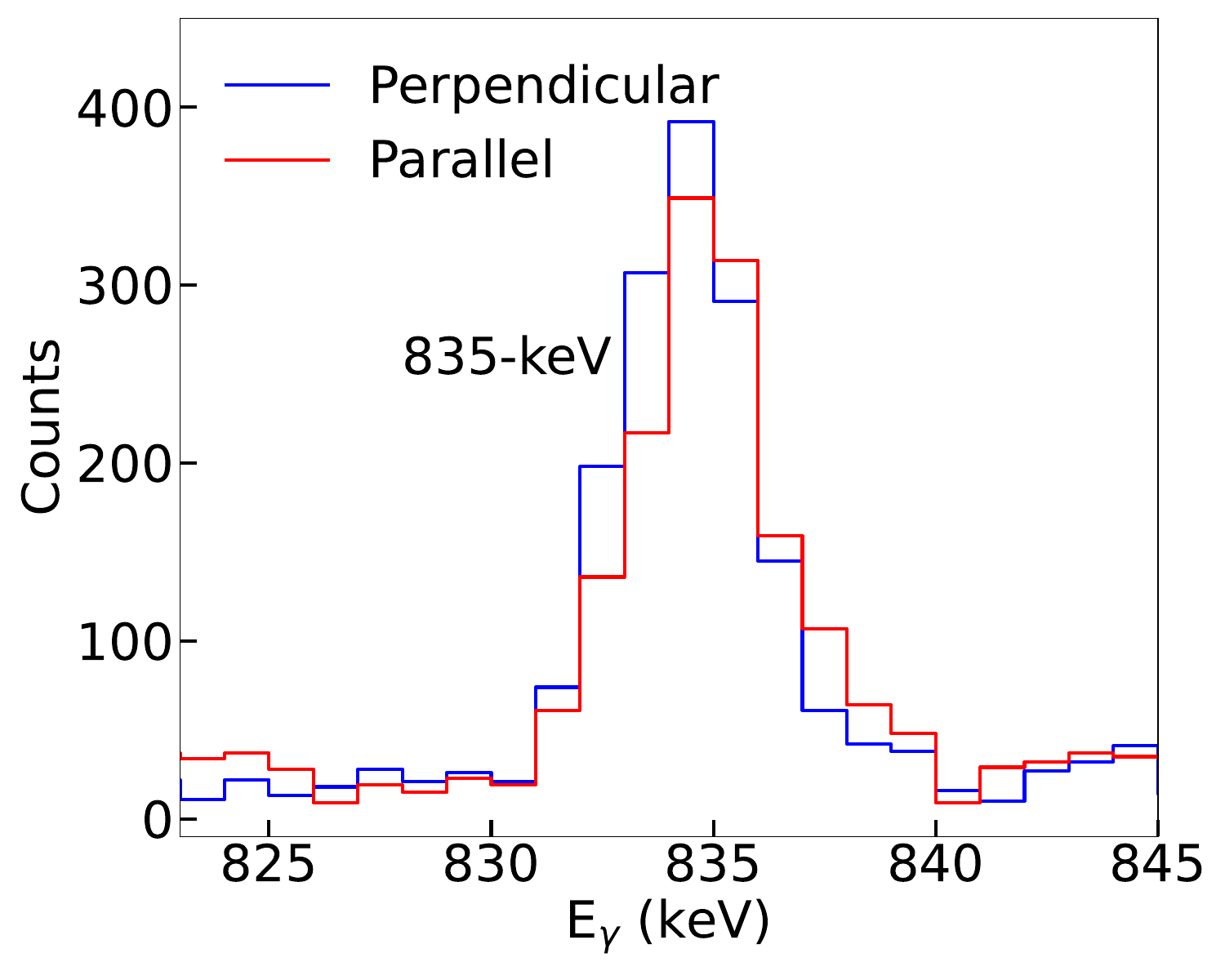}
\caption{\label{fig6}Spectra of 835-keV transition peak corresponding to the perpendicular and the parallel scattering events in the HPGe clover detectors at 90$^o$.}
\end{figure}

\begin{figure}
\includegraphics[angle=0,scale=.27,trim=0.0cm 0.0cm 0.0cm 0.0cm,clip=true]{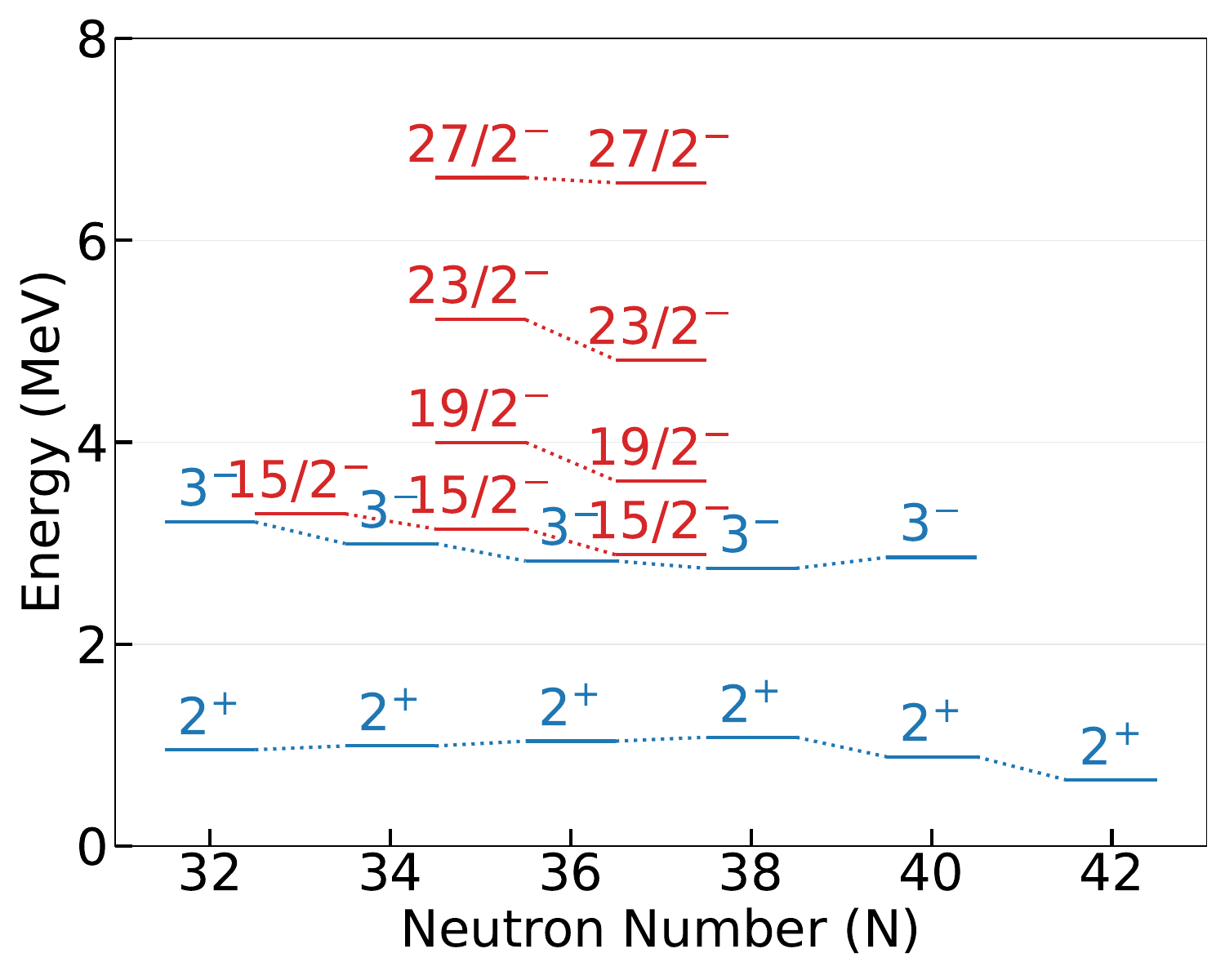}
\caption{\label{fig7}Evolution of excitation energy of the 3$^-$ state across the even-A Zn isotopes. The states of the octupole sequence, observed in odd-A Zn isotopes, are plotted along with; 
the excitation energies have been plotted w.r.t that of the respective 9/2$^+$ state.}
\end{figure}

\section{Discussions}

One of the principal impetuses for nuclear structure pursuits around magicities 
is the validation of shell model calculations or, more specifically, the residual interactions
proposed for the respective model space. Spectroscopic data provide an index not only for testing the updated interaction 
Hamiltonians but also for investigating the role of different orbitals in the level structure of the nucleus and/ or the 
efficacy of different truncation schemes in the interpretation of the underlying excitations. The recent studies 
on nuclei around \cite{Aya22} the closure at $N, Z = 28$, or even on those with considerable number nucleons 
outside the closure \cite{Aya23,Sha26}, have used large basis shell model calculations, to interpret the level structures 
at least up to low/ moderate excitations. In $^{66}$Zn \cite{Aya22}, for example, the overlap between the calculated and the experimental level energies 
is reasonable up to $\approx$ 10$^+$ state but become increasingly deviant at higher spins. It is expected that more of the improved data sets, to compare with the 
calculations, shall facilitate in better constraining the model parameters and help in the formulation of
interactions that would translate into faithful microscopic description of the respective level schemes. \\

\LTcapwidth=\textwidth
\begin{longtable*}{cccccccccccccccc}
\caption{\label{tab1} Experimental and calculated level energies of $^{65}$Zn along with the average particle occupancies of the states.
The calculations have been carried out using two different interactions, JUN45 and JJ44B, and the respective results are accordingly recorded in the table.} \\
\hline
Exp       & $J^{\pi}$ & JUN45       & & $p_{3/2}$ & $f_{5/2}$ & $p_{1/2}$ & $g_{9/2}$ &    & JJ44B & & $p_{3/2}$ & $f_{5/2}$ & $p_{1/2}$ & $g_{9/2}$ &  \\
(keV)     &           &  (keV)      & &           &           &           &           &    & (keV) & &           &           &           &           &  \\
\hline
\endfirsthead

\multicolumn{16}{c}{{\tablename\ \thetable{} -- continued from previous page}} \\
\hline
Exp (keV) & $J^{\pi}$ & JUN45 (keV) & & $p_{3/2}$ & $f_{5/2}$ & $p_{1/2}$ & $g_{9/2}$ & \% & JJ44B & & $p_{3/2}$ & $f_{5/2}$ & $p_{1/2}$ & $g_{9/2}$ &  \\
(keV)     &           &  (keV)      & &           &           &           &           &    & (keV) & &           &           &           &           &  \\
\hline
\endhead

\hline \multicolumn{16}{c}{{continued on next page}} \\
\endfoot

\hline
\endlastfoot

          &                   &      &   &      &      &      &      &      &      &   &      &      &      &      &    \\
    \multicolumn{14}{c}{\bf{Negative Parity States}} \\
    0     & 5/2$^{-}_{1}$     & 0     & p & 1.54   & 0.14   & 0.24   & 0.09   &      & 266    & p & 0.36  & 1.10   & 0.36   & 0.17 &    \\
          &                   &       & n & 3.12   & 2.85   & 0.60   & 0.43   &      &        & n & 2.86  & 2.97   & 0.57   & 0.59 &  \\
    54    & 1/2$^{-}_{1}$     & 525   & p & 1.58   & 0.15   & 0.18   & 0.09   &      & 339    & p & 0.26  & 1.29   & 0.27   & 0.18 &    \\
          &                   &       & n & 3.11   & 2.43   & 0.98   & 0.47   &      &        & n & 2.35  & 3.01   & 1.06   & 0.57 &  \\
    115   & 3/2$^{-}_{1}$     & 173   & p & 1.44   & 0.14   & 0.34   & 0.08   &      & 0      & p & 0.36  & 1.07   & 0.40   & 0.16 &    \\
          &                   &       & n & 3.04   & 3.03   & 0.53   & 0.39   &      &        & n & 3.01  & 2.96   & 0.50   & 0.54 &  \\
    207   & 3/2$^{-}_{2}$     & 377   & p & 1.52   & 0.15   & 0.25   & 0.08   &      & 697    & p & 0.42  & 1.15   & 0.27   & 0.16 &    \\
          &                   &       & n & 2.88   & 3.01   & 0.68   & 0.43   &      &        & n & 2.58  & 2.70   & 0.90   & 0.82 &  \\
    769   & 5/2$^{-}_{2}$     & 950   & p & 1.46   & 0.16   & 0.31   & 0.07   &      & 1093   & p & 0.30  & 1.24   & 0.29   & 0.17 &    \\
          &                   &       & n & 2.65   & 3.30   & 0.66   & 0.38   &      &        & n & 2.49  & 2.93   & 1.06   & 0.52 &  \\
    864   & 7/2$^{-}_{1}$     & 919   & p & 1.48   & 0.16   & 0.29   & 0.07   &      & 1092   & p & 0.47  &  0.98  & 0.42   & 0.12 &    \\
          &                   &       & n & 3.09   & 2.87   & 0.68   & 0.36   &      &        & n & 3.00  & 2.95   & 0.55   & 0.51 &  \\
    1252  & 7/2$^{-}_{2}$     & 1671  & p & 1.52   & 0.18   & 0.23   & 0.08   &      & 1897   & p & 0.45  & 1.15   & 0.25   & 0.15 &    \\
          &                   &       & n & 2.69   & 2.90   & 0.95   & 0.46   &      &        & n & 2.58  & 2.61   & 0.98   & 0.83 &  \\
    1263  & 9/2$^{-}_{1}$     & 1179  & p & 1.49   & 0.16   & 0.28   & 0.07   &      & 1551   & p & 0.50  & 1.00   & 0.37   & 0.14 &    \\
          &                   &       & n & 3.12   & 2.93   & 0.58   & 0.37   &      &        & n & 2.92  & 2.92   & 0.56   & 0.59 &  \\
    1908  & 9/2$^{-}_{2}$     & 2094  & p & 1.58   & 0.16   & 0.19   & 0.08   &      & 2391   & p & 0.41  & 1.15   & 0.30   & 0.14 &   \\
          &                   &       & n & 3.06   & 2.84   & 0.77   & 0.32   &      &        & n & 2.41  & 2.91   & 1.17   & 0.51 &  \\
    2303  & 11/2$^{-}_{1}$    & 2435  & p & 1.42   & 0.22   & 0.29   & 0.07   &      & 2472   & p & 0.86  & 0.82   & 0.21   & 0.11 &   \\
          &                   &       & n & 2.85   & 3.18   & 0.65   & 0.32   &      &        & n & 3.02  & 2.81   & 0.57   & 0.60 &  \\
    2399  & 11/2$^{-}_{2}$    & 2676  & p & 1.45   & 0.23   & 0.25   & 0.07   &      & 3184   & p & 0.89  & 0.79   & 0.25   & 0.07 &   \\
          &                   &       & n & 2.83   & 3.22   & 0.63   & 0.32   &      &        & n & 2.77  & 2.21   & 0.62   & 1.40 &  \\
    2930  & 13/2$^{-}_{1}$    & 3008  & p & 1.30   & 0.37   & 0.27   & 0.06   &      & 3114   & p & 0.89  & 0.84   & 0.18   & 0.10 &   \\
          &                   &       & n & 3.08   & 2.95   & 0.59   & 0.38   &      &        & n & 2.80  & 2.87   & 0.62   & 0.71 &  \\
    4207  & 15/2$^{-}_{1}$    & 4670  & p & 1.06   & 0.47   & 0.39   & 0.09   &      & 3859   & p & 0.68  & 0.76   & 0.43   & 0.14 &   \\
          &                   &       & n & 2.31   & 2.33   & 0.43   & 1.92   &      &        & n & 2.47  & 2.02   & 0.49   & 2.02 &  \\
    4254  & 15/2$^{-}_{2}$    & 4754  & p & 1.05   & 0.70   & 0.20   & 0.06   &      & 4450   & p & 0.98  & 0.83   & 0.10   & 0.09 &   \\
          &                   &       & n & 2.80   & 2.93   & 0.70   & 0.56   &      &        & n & 3.12  & 2.70   & 0.69   & 0.49 &  \\
    4865  & 17/2$^{-}_{1}$    & 4861  & p & 1.06   & 0.46   & 0.41   & 0.06   &      & 4217   & p & 0.77  & 0.70   & 0.44   & 0.10 &   \\
          &                   &       & n & 2.22   & 2.25   & 0.45   & 2.08   &      &        & n & 2.56  & 1.87   & 0.49   & 2.07 &  \\
    4892  & 17/2$^{-}_{2}$    & 4999  & p & 0.93   & 1.02   & 0.02   & 0.04   &      & 4879   & p & 0.71  & 0.75   & 0.42   & 0.12 &   \\
          &                   &       & n & 3.08   & 2.74   & 0.78   & 0.40   &      &        & n & 2.16  &  2.27  & 0.53   & 2.04 &  \\
    5065  & 19/2$^{-}_{1}$    & 5203  & p & 1.05   & 0.44   & 0.42   & 0.10   &      & 4518   & p & 0.72  & 0.69   & 0.42   & 0.17 &   \\
          &                   &       & n & 2.11   & 2.44   & 0.42   & 2.03   &      &        & n & 2.64  & 1.91   & 0.46   & 1.98 &   \\
    5411  & 21/2$^{-}_{1}$    & 5319  & p & 1.09   & 0.45   & 0.40   & 0.07   &      & 5043   & p & 0.78  & 0.71   & 0.41   & 0.09 &   \\
          &                   &       & n & 2.31   & 2.19   & 0.41   & 2.09   &      &        & n & 2.43  & 1.98   & 0.51   & 2.08 &  \\
    5668  & 21/2$^{-}_{2}$    & 6232  & p & 1.11   & 0.47   & 0.32   & 0.10   &      & 5658   & p & 0.81  & 0.65   & 0.28   & 0.27 &   \\
          &                   &       & n & 1.97   & 2.44   & 0.54   & 2.05   &      &        & n & 1.96  & 2.42   & 0.67   & 1.95 &  \\
    5770  & 23/2$^{-}_{1}$    & 6209  & p & 0.94   & 0.58   & 0.38   & 0.10   &      & 5530   & p & 0.85  & 0.50   & 0.29   & 0.36 &   \\
          &                   &       & n & 2.12   & 2.42   & 0.43   & 2.03   &      &        & n & 2.74  & 1.96   & 0.49   & 1.81 &  \\
    6280  & 23/2$^{-}_{2}$    & 7064  & p & 1.04   & 0.49   & 0.36   & 0.11   &      & 6057   & p & 0.96  & 0.17   & 0.09   & 0.77 &   \\
          &                   &       & n & 2.14   & 2.41   & 0.42   & 2.04   &      &        & n & 2.47  & 2.39   & 0.76   & 1.39 &  \\
    6842  & 25/2$^{-}_{1}$    & 6386  & p & 0.99   & 0.56   & 0.39   & 0.05   &      & 6335   & p & 0.89  & 0.68   & 0.36   & 0.08 &   \\
          &                   &       & n & 2.23   & 2.28   & 0.41   & 2.08   &      &        & n & 2.39  & 2.01   & 0.52   & 2.08 &  \\
	  &                   &       &   &        &        &        &        &      &        &   &       &        &        &      &    \\
           \multicolumn{14}{c}{\bf{Positive Parity States}} \\
          &                   &       &   &        &        &        &        &      &        &   &       &        &        &      &    \\ 
	  \hline
     1066 & 9/2$^{+}_{1}$     & 1066 & p  & 1.31   & 0.26   & 0.36   & 0.07   &      & 743    & p & 0.57  & 0.87   & 0.44   & 0.11 &       \\
          &                   &      & n  & 2.61   & 2.59   & 0.54   & 1.26   &      &        & n & 2.68  & 2.49   & 0.52   & 1.31 &     \\
     2054 & 13/2$^{+}_{1}$    & 1814 & p  & 1.21   & 0.31   & 0.41   & 0.06   &      & 1503   & p & 0.62  & 0.81   & 0.48   & 0.09 &       \\
          &                   &      & n  & 2.54   & 2.69   & 0.54   & 1.23   &      &        & n & 2.78  &  2.45  & 0.50   & 1.28 &     \\
     2137 & 11/2$^{+}_{1}$    & 2462 & p  & 1.31   & 0.25   & 0.37   & 0.07   &      & 2343   & p & 0.56  & 0.93   & 0.39   & 0.12 &       \\
          &                   &      & n  & 2.55   & 2.71   & 0.52   & 1.22   &      &        & n & 2.55  & 2.57   & 0.64   & 1.24 &     \\
     2923 & 13/2$^{+}_{2}$    & 3068 & p  & 1.34   & 0.27   & 0.31   & 0.07   &      & 2934   & p & 0.50  & 0.93   & 0.42   & 0.16 &       \\
          &                   &      & n  & 2.76   & 2.56   & 0.51   & 1.16   &      &        & n & 2.57  &  2.71  & 0.57   & 1.15 &     \\
     3227 & 17/2$^{+}_{1}$    & 3015 & p  & 1.12   & 0.44   & 0.37   & 0.06   &      & 2657   & p & 0.79  & 0.77   & 0.38   & 0.07 &       \\
          &                   &      & n  & 2.64   & 2.63   & 0.52   & 1.21   &      &        & n & 2.76  & 2.46   & 0.52   & 1.26 &     \\
     3375 & 13/2$^{+}_{3}$    & 3397 & p  & 1.30   & 0.29   & 0.35   & 0.07   &      & 3242   & p & 0.60  & 0.87   & 0.43   & 0.10 &        \\
          &                   &      & n  & 2.69   & 2.59   & 0.55   & 1.17   &      &        & n & 2.74  & 2.50   & 0.58   & 1.18 &     \\
     3472 & 15/2$^{+}_{1}$    & 3337 & p  & 1.38   & 0.27   & 0.27   & 0.08   &      & 3486   & p & 0.68  & 0.82   & 0.40   & 0.10 &       \\
          &                   &      & n  & 2.83   & 2.51   & 0.50   & 1.16   &      &        & n & 2.83  &  2.41  & 0.51   & 1.24 &     \\
     3711 & 15/2$^{+}_{2}$    & 3506 & p  & 1.25   & 0.30   & 0.38   & 0.07   &      & 3705   & p & 0.60  & 0.88   & 0.38   & 0.14 &       \\
          &                   &      & n  & 2.57   & 2.79   & 0.47   & 1.18   &      &        & n & 2.47  & 2.79   & 0.53   & 1.21 &     \\
     3784 & 17/2$^{+}_{2}$    & 3718 & p  & 1.40   & 0.28   & 0.25   & 0.07   &      & 3765   & p & 1.01  & 0.78   & 0.08   & 0.14 &       \\
          &                   &      & n  & 2.79   & 2.42   & 0.60   & 1.19   &      &        & n & 2.34  & 2.69   & 0.76   & 1.20 &     \\
     4078 & 17/2$^{+}_{3}$    & 4169 & p  & 1.22   & 0.39   & 0.33   & 0.06   &      & 4075   & p & 0.74  & 0.86   & 0.28   & 0.13 &       \\
          &                   &      & n  & 2.70   & 2.63   & 0.51   & 1.16   &      &        & n & 2.74  & 2.50   & 0.55   & 1.21 &     \\
     4623 & 19/2$^{+}_{1}$    & 4801 & p  & 1.20   & 0.42   & 0.32   & 0.05   &      & 4835   & p & 0.87  & 0.75   & 0.29   & 0.09 &       \\
          &                   &      & n  & 2.69   & 2.63   & 0.52   & 1.17   &      &        & n & 2.81  &  2.45  & 0.53   & 1.21 &     \\
     4880 & 19/2$^{+}_{2}$    & 4904 & p  & 1.18   & 0.42   & 0.34   & 0.06   &      & 5031   & p & 0.97  & 0.79   & 0.12   & 0.13 &       \\
          &                   &      & n  & 2.77   & 2.60   & 0.47   & 1.16   &      &        & n & 2.48  & 2.53   & 0.77   & 1.22 &     \\
     4934 & 21/2$^{+}_{1}$    & 4618 & p  & 1.01   & 0.64   & 0.30   & 0.05   &      & 4266   & p & 0.91  & 0.76   & 0.27   & 0.06 &       \\
          &                   &      & n  & 2.72   & 2.56   & 0.53   & 1.20   &      &        & n & 2.71  & 2.52   & 0.55   & 1.22 &     \\
     5492 & 21/2$^{+}_{2}$    & 5351 & p  & 1.09   & 0.59   & 0.28   & 0.05   &      & 5584   & p & 0.90  &  0.80  & 0.23   & 0.07 &       \\
          &                   &      & n  & 2.72   & 2.48   & 0.60   & 1.21   &      &        & n & 2.82  & 2.38   & 0.59   & 1.20 &     \\
     6142 & 23/2$^{+}_{1}$    & 6066 & p  & 0.89   & 1.05   & 0.02   & 0.04   &      & 6259   & p & 1.07  & 0.85   & 0.02   & 0.06 &       \\
          &                   &      & n  & 2.91   & 2.17   & 0.72   & 1.20   &      &        & n & 2.08  & 2.79   & 0.93   & 1.20 &     \\
     6688 & 25/2$^{+}_{1}$    & 6297 & p  & 0.89   & 1.01   & 0.06   & 0.04   &      & 6165   & p & 1.04  & 0.82   & 0.08   & 0.06 &       \\
          &                   &      & n  & 3.10   & 2.30   & 0.41   & 1.19   &      &        & n & 2.45  & 2.71   & 0.66   & 1.19 &     \\
\end{longtable*}

The $^{65}$Zn nucleus has two protons ($\pi$) and seven neutrons ($\nu$) outside the doubly magic $^{56}$Ni-core.
The valence nucleons are distributed in the model space of $p_{3/2}, f_{5/2}, p_{1/2}, g_{9/2}$ orbitals.
The present calculations in the framework of the large basis shell model have been implemented using the KSHELL \cite{Shi19} code and 
are based on two interactions, JUN45 \cite{Hon09} and JJ44B \cite{Lis04} that
are of widespread use \cite{Aya22,Aya23,Sha26} in this mass region. Fig.~8 illustrates the comparison between the experimental and the theoretical
level energies, corresponding to the two interactions. Table~2 records the experimental and the theoretical level energies along with the 
respective average particle occupancies for the states. The energies of the negative parity states, in general, are better reproduced 
in the calculation using the JUN45 interaction compared to that using the JJ44B. Importantly, the JJ44B Hamiltonian could not 
reproduce the 5/2$^-$ level as the ground state of the $^{65}$Zn nucleus while the use of the JUN45 interaction results in the
correct ground state. Further, in the calculations with JUN45, the yrast negative parity states are closer (within $\lessapprox$ 100-150 keV) 
to the experimental values; this, however, does not hold for 1/2$^-$, 15/2$^-$ and the highest negative parity levels that substantially ($\gtrapprox$ 500 keV) 
differ in their theoretical and measured values.
Similarly, the non-yrast (negative parity) states for JUN45, except for the 17/2$^-$, are only in modest agreement (difference $\gtrapprox$ 200-400 keV)
with the measurements. The large differences between the calculated and the experimental level energies also characterize the results of the
JJ44B interaction, except for the non-yrast 17/2$^-$ and 21/2$^-$ states. But these are rather limited to establish the merit of the 
interaction in the present context. As far as the particle configurations calculated by the JUN45 interaction is concerned, the lower (upto 13/2$^-$) negative parity states 
are dominated by the occupancies $\pi(p_{3/2}f_{5/2}p_{1/2})^2\otimes\nu(p_{3/2}f_{5/2})^6p_{1/2}^1$ and $\pi(p_{3/2}f_{5/2}p_{1/2})^2\otimes\nu(p_{3/2}f_{5/2})^6g_{9/2}^1$.
For the negative parity states of higher ($\gtrapprox$ 15/2) spins, the configuration majorly assumes the occupancies $\pi(p_{3/2}f_{5/2}p_{1/2})^2\otimes\nu(p_{3/2}f_{5/2})^5g_{9/2}^2$.
The proton occupancies individually evolve from $\pi p_{3/2}^2$ for lower (upto 13/2$^-$) levels to $\pi p_{3/2}^1(f_{5/2}p_{1/2})^1$ for the higher ones.
The differences in the level energy results of the JUN45 and the JJ44B interactions also show up in the respective particle configurations.
The major proton occupancies corresponding to the JJ44B interaction are $\pi(p_{3/2}p_{1/2})^1f_{5/2}^1$ for the lower (upto 9/2$^-$) negative parity 
states and $\pi p_{3/2}^1f_{5/2}^1$ for most of the higher ones. The major neutron occupancies are close to those resulting from the use of the JUN45 interaction,
$\nu(p_{3/2}f_{5/2})^6p_{1/2}^{0,1}g_{9/2}^{0,1}$ for the lower spin ($\lessapprox$13/2$^-$) states and $\nu(p_{3/2}f_{5/2}p_{1/2})^5g_{9/2}^{2}$ for the higher 
ones, except 15/2$^-_{2}$ and 23/2$^-_{2}$. \\

\begin{figure*}
\includegraphics[angle=0,scale=.30,trim=0.0cm 0.0cm 0.0cm 0.0cm,clip=true]{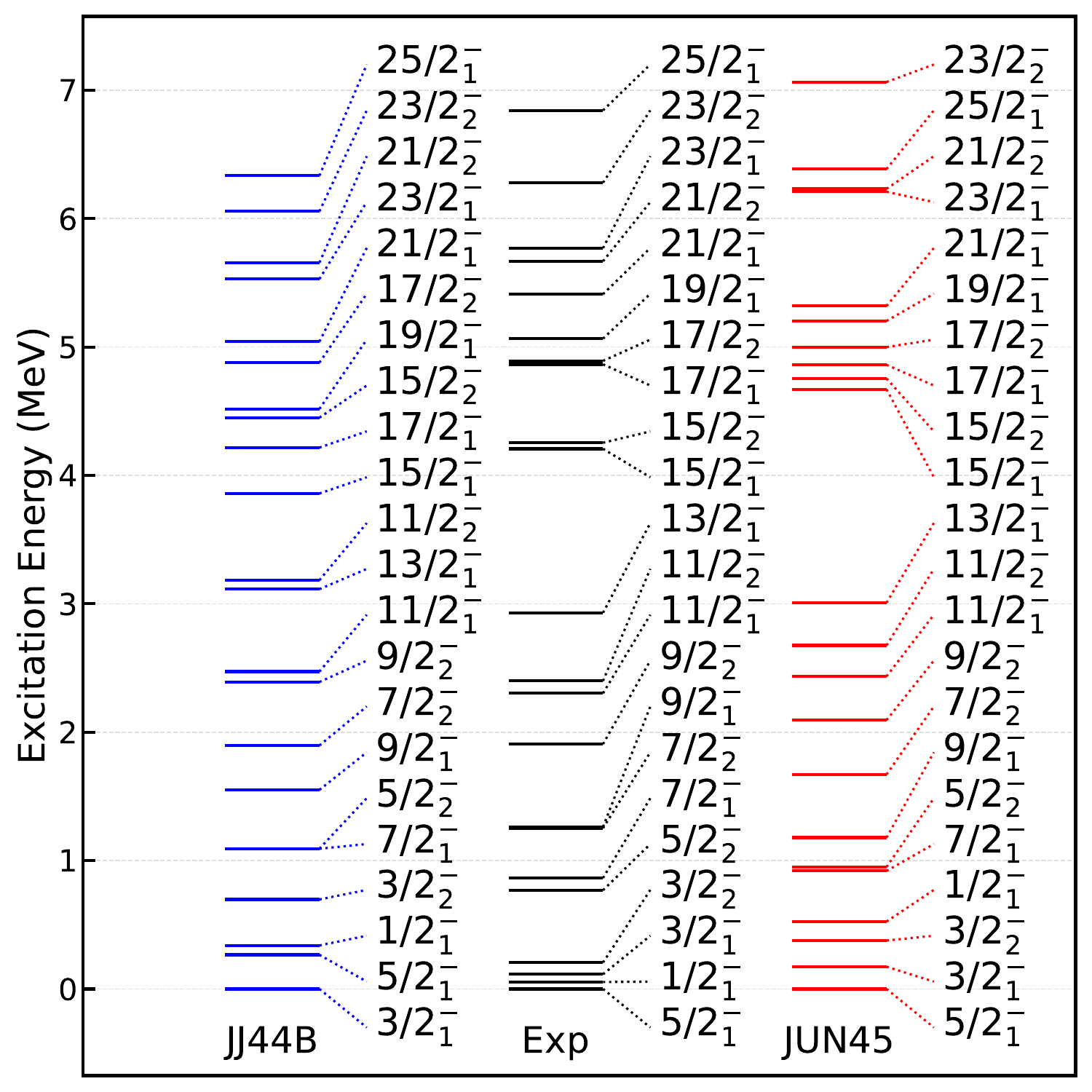}
\includegraphics[angle=0,scale=.30,trim=0.0cm 0.0cm 0.0cm 0.0cm,clip=true]{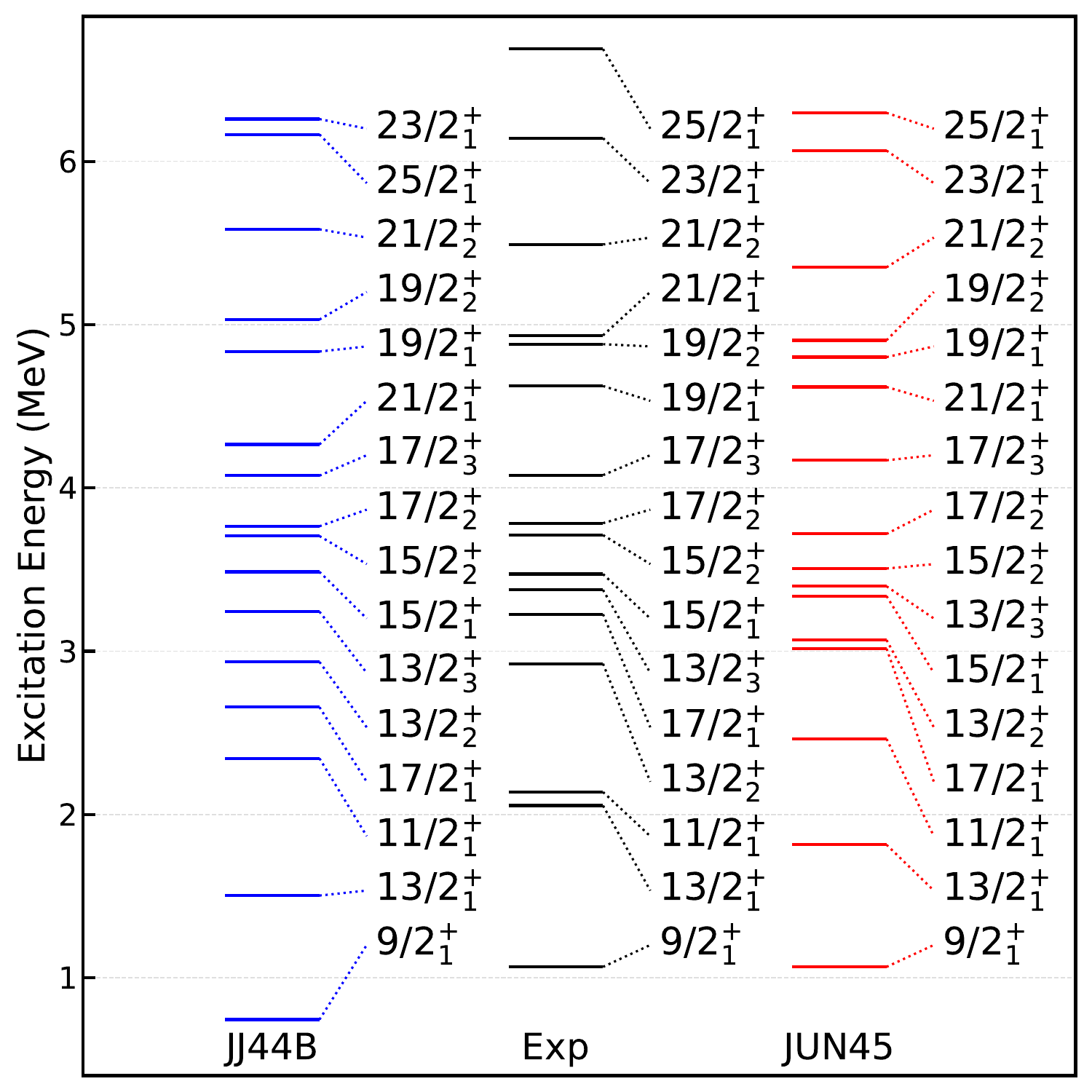}
\caption{\label{fig8}Comparison between experimental and shell model calculated level energies using JUN45 and JJ44B interactions.}
\end{figure*}

\begin{figure*}
\includegraphics[angle=0,scale=.30,trim=0.0cm 0.0cm 0.0cm 0.0cm,clip=true]{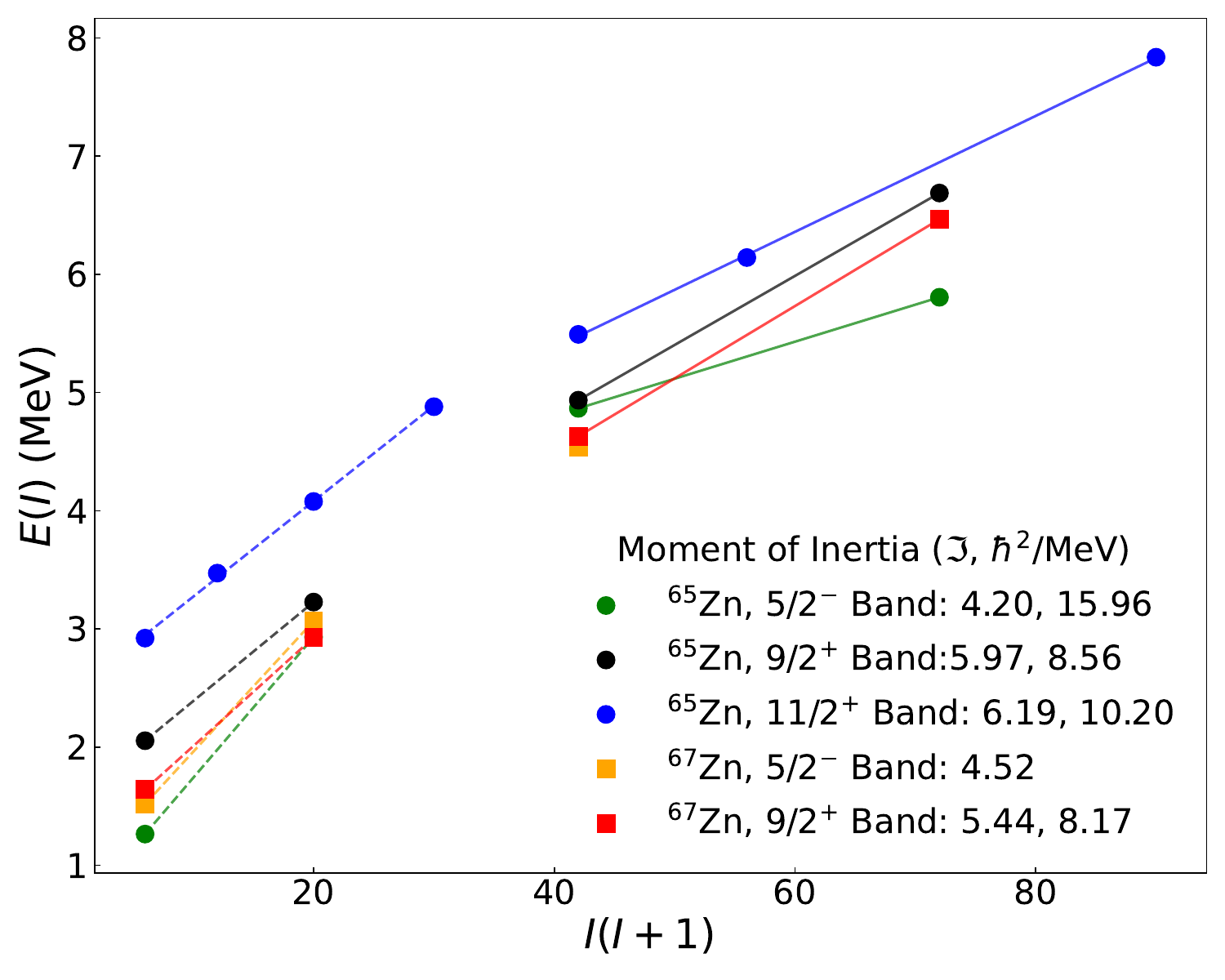}
\includegraphics[angle=0,scale=.30,trim=0.0cm 0.0cm 0.0cm 0.0cm,clip=true]{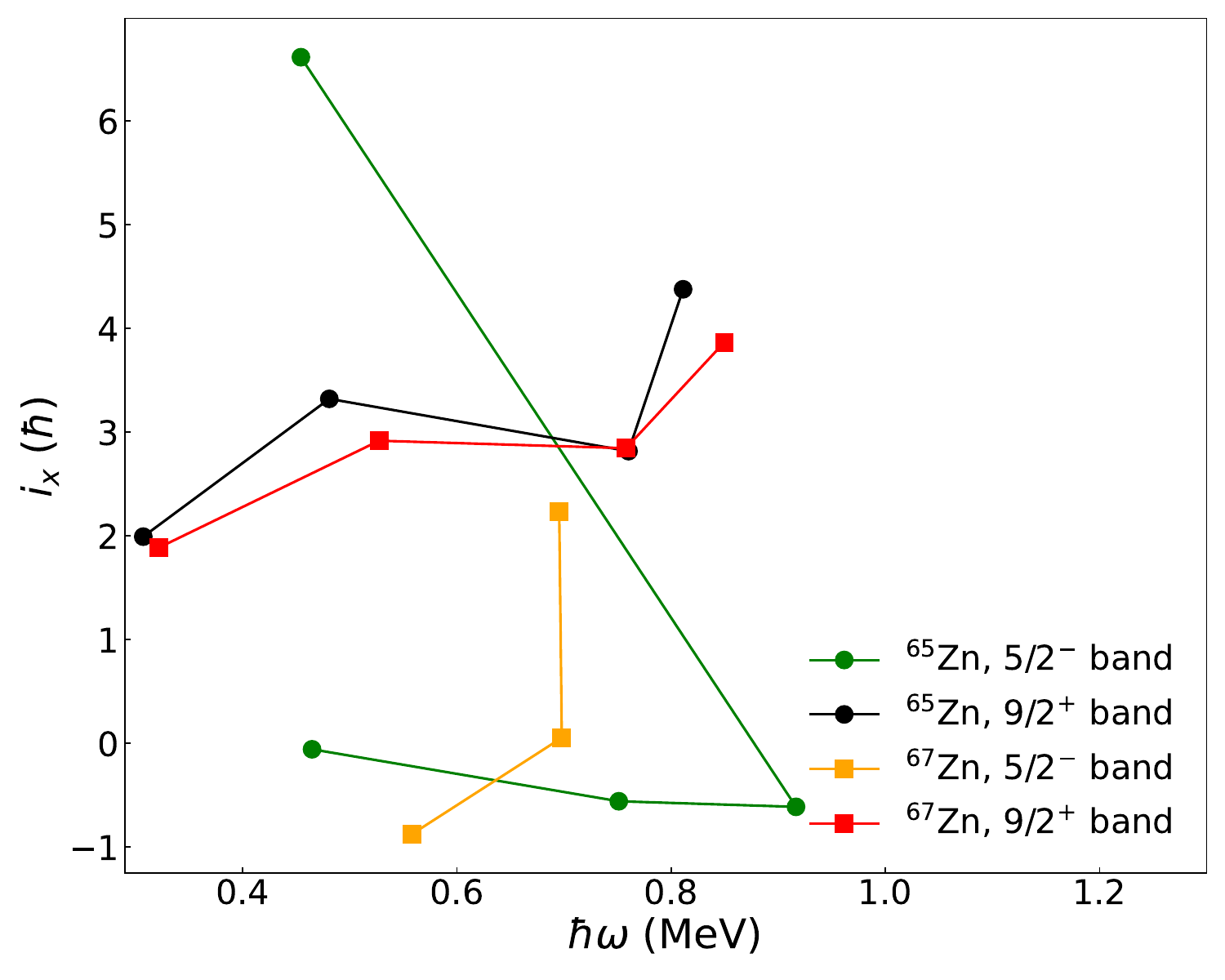}
\caption{\label{fig9}(a) Plot of excitation energy (E(I)) versus spin (I(I+1)) for rotational bands of the $^{65}$Zn and $^{67}$Zn isotopes. The fitted values of the 
moment of inertia, before and after alignment in each band, are stated in the inset. (b) Plot of the aligned angular momentum ($i_x$) versus rotational frequency ($\hbar\omega$) for the
5/2$^-$ and 9/2$^+$ bands of $^{65}$Zn and $^{67}$Zn. The equations and the parameters used for the calculations are detailed in the text.}
\end{figure*}

It is somewhat difficult to generalize on the overlap between the calculated and the experimental level energies, for the
positive parity states of $^{65}$Zn. As recorded in Table~2, there are scattered instances of superior agreement for both 
the calculations. The lowest positive parity state at 9/2$^+$ is exactly reproduced in the calculation with the JUN45 interaction
but not in that with the JJ44B. The theoretical values for the non-yrast 13/2$^+_3$, 17/2$^+_2$, 19/2$^+_2$ and the yrast 21/2$^+_1$, as calculated using the  
JUN45 Hamiltonian, are in superlative ($\lessapprox$ 50 keV) to reasonable ($\lesssim$ 100 keV) compliance with their measured values.
Similarly, the calculated energies for the non-yrast 13/2$^+_2$, for both the 15/2$^+$, and for the non-yrast 17/2$^+_3$, corresponding to the
JJ44B interaction, overlap very well (within $\lessapprox$ 20 keV) with the respective experimental numbers. 
The other states, however, are only modestly represented in both the calculations, and the perspectives on the success of the
shell model calculations, for calculating the positive parity states of $^{65}$Zn, remain largely inconclusive.
Just as the negative parity states, the particle configurations for the positive parity levels differ in the two calculations.
The results of the JUN45 interaction indicate the dominant occupancies to be $\pi p_{3/2}^1(f_{5/2}p_{1/2})^1 \otimes \nu(p_{3/2}f_{5/2}p_{1/2})^6g_{9/2}^1$.
The calculations with the JJ44B interaction point to different proton occupancies, $\pi f_{5/2}^1(p_{3/2}p_{1/2})^1$ for the lower excited states ($\lessapprox$ 15/2$^+$)
and $\pi f_{5/2}^1p_{3/2}^1$ for the higher ones. It might still require additional data, on other nuclei in this region, to further constrain the calculations 
and facilitate the development of model interactions that would better represent their level structures in the microscopic paradigm. \\

\begin{figure}
\includegraphics[angle=0,scale=.35,trim=8.0cm 3.0cm 0.0cm 3.0cm,clip=true]{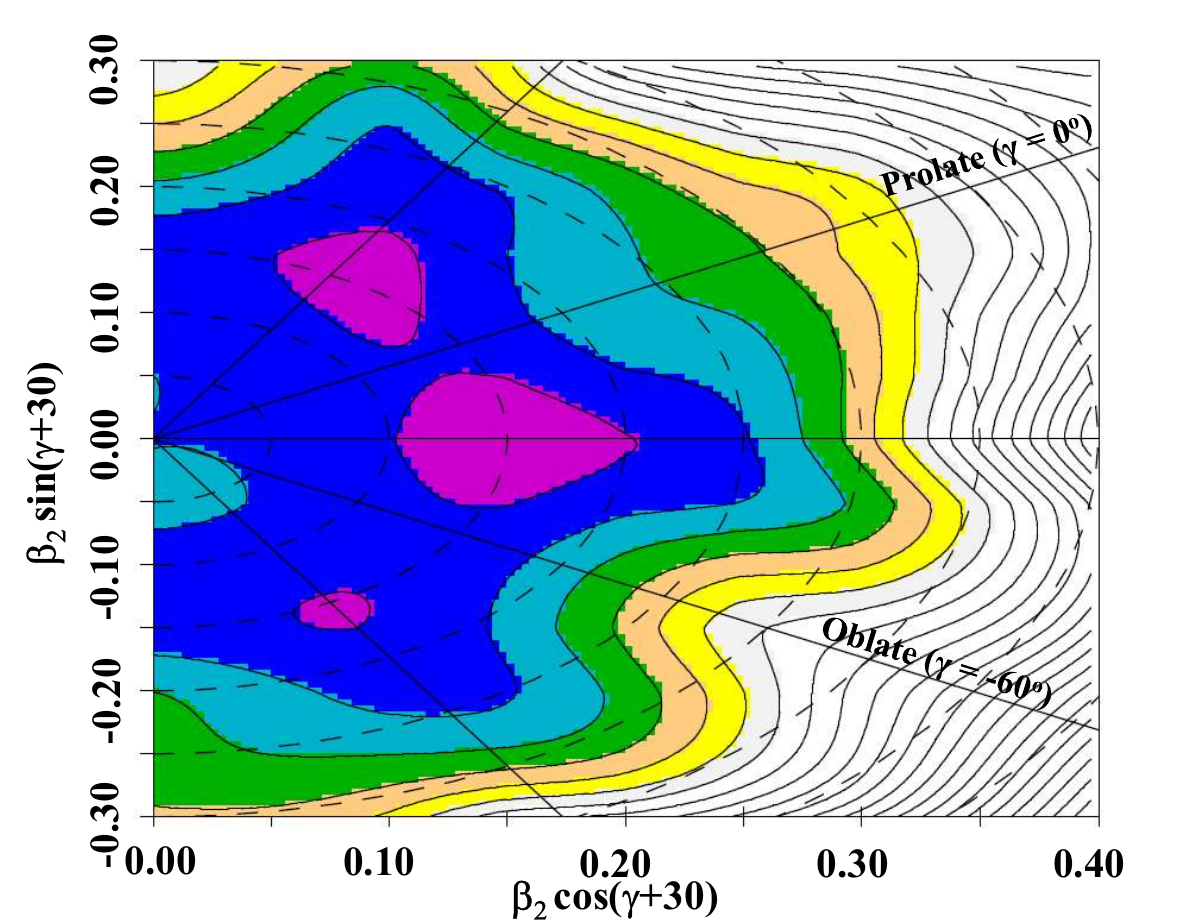}
\caption{\label{fig10}TRS plot of the 5/2$^-$ ground state band of the $^{65}$Zn nucleus. This plot corresponds to the minimum rotational frequency ($\hbar\omega$), near the bandhead. The 
potential minimum is at $\beta_2$ $\approx$ 0.15 with considerable degree of $\gamma$-softness.}
\end{figure}

\begin{figure}
\includegraphics[angle=0,scale=.35,trim=6.0cm 3.0cm 0.0cm 3.0cm,clip=true]{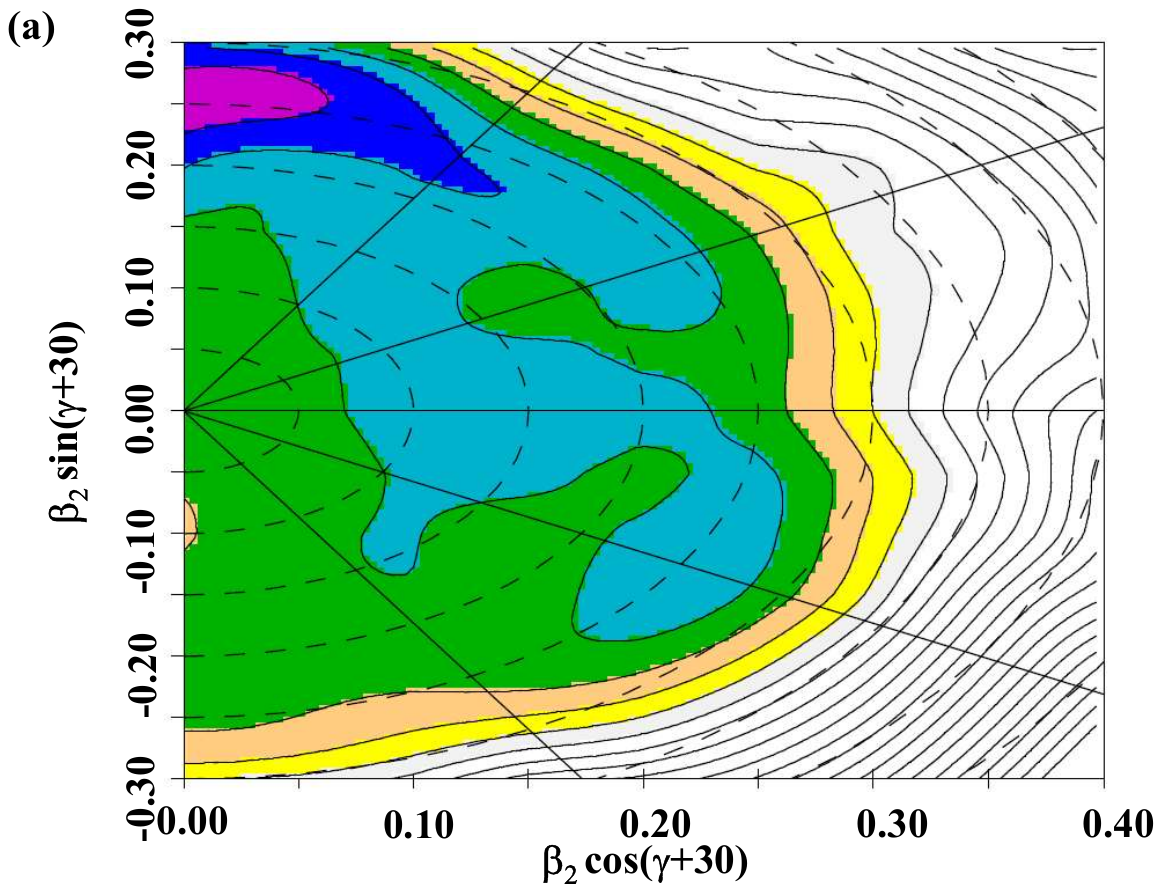}
\includegraphics[angle=0,scale=.35,trim=6.0cm 3.0cm 0.0cm 3.0cm,clip=true]{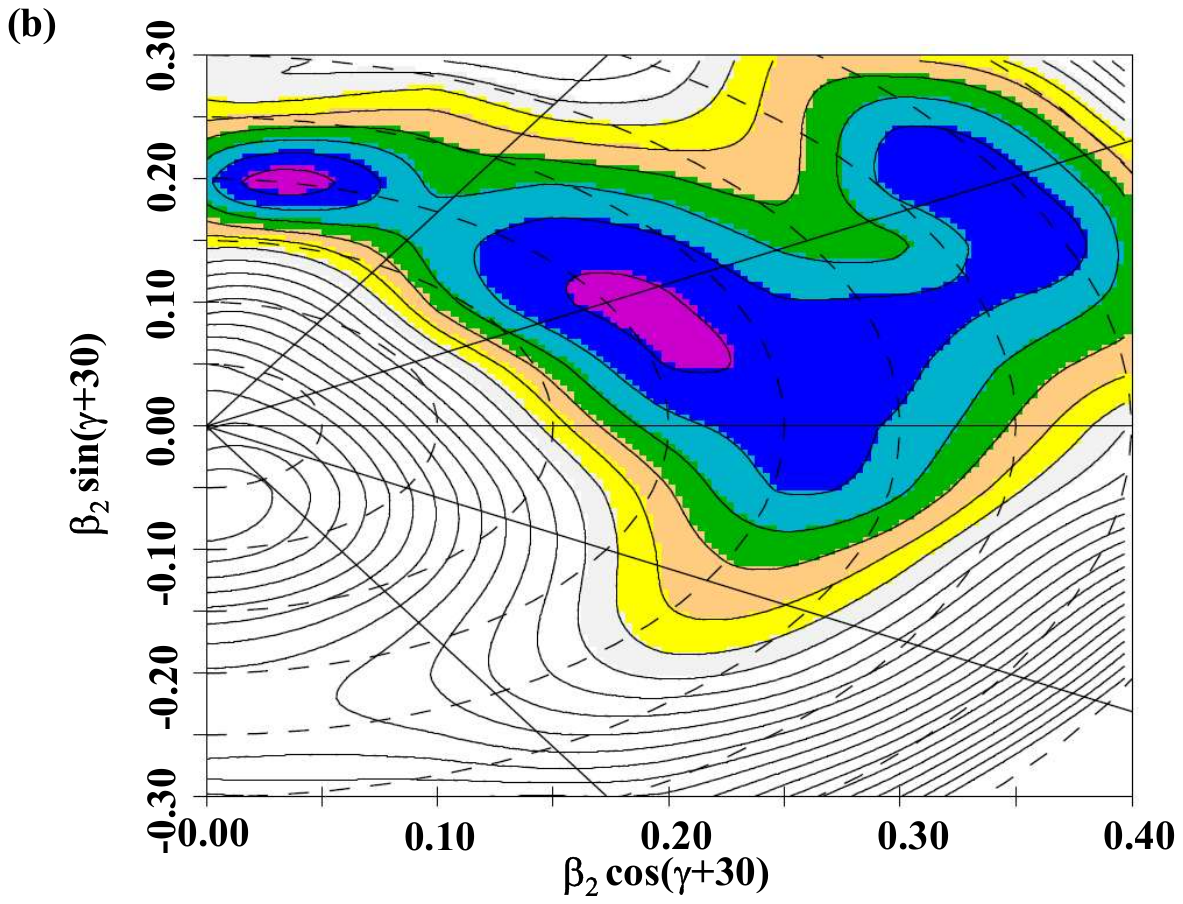}
\caption{\label{fig11}TRS plot of the 9/2$^+$ band of the $^{65}$Zn nucleus. The plots are for (a) the minimum rotational frequency ($\hbar\omega$), near the bandhead and 
(b) $\hbar\omega$ = 0.6 corresponding to the 17/2$^+$ state of the band.}
\end{figure}

Further to the shell model calculations, it is worth discussing the deformation characteristics of the $^{65}$Zn nucleus 
particularly in the context of the considerable number (two protons, seven neutrons) of valence nucleons outside the doubly-magic Ni-core and the 
neutron occupancy of the deformation-driving $g_{9/2}$ orbital. Of particular interest in this context are the bands B1-B3 that
are envisaged to embody distinct features of collectivity and shape characteristics. Fig.~9a illustrates the 
level excitation energy versus spin data, for the rotational bands of $^{65}$Zn and $^{67}$Zn isotopes. 
The data of $^{67}$Zn are adopted from Ref.~\cite{Zha25}.
The $E(I)$ versus $I(I+1)$ data of each band
have been fitted individually over regions before and after the alignment, observed therein (discussed hereafter). The corresponding function is 
$E_I = E_0 + (\hbar^2/2\Im)I(I+1)$ and the respective values of the moment of inertia (MOI, $\Im$) are stated in the inset of the plot (Fig.~9(a)).
The MOI associated with the first part of the ground state 5/2$^-$ band is similar for $^{65}$Zn, studied in this work, and $^{67}$Zn, last studied by Zhang {\it{et al.}} \cite{Zha25}. 
The present investigation has extended the band in $^{65}$Zn 
to higher excitations along with modified spin-parity assignments therein. The plot of aligned angular momentum ($i_x$) against rotational frequency ($\hbar\omega$) 
for the bands in $^{65}$Zn (and $^{67}$Zn) are illustrated in Fig.~9(b).
The aligned angular momentum is calculated using \cite{Reg03} $i_x = I_x - I_{ref}$ where $I_x = \sqrt{I(I+1) - K^2}$ and $I_{ref} = (J_0 + \omega^2J_1)\omega$; $J_0$ = 6.0 $\hbar^2/MeV$ and $J_1$ = 3.5 $\hbar^4/MeV^3$ 
are Harris parameters adopted from Ref.~\cite{Rai20}.
As far as the 5/2$^-$ band is concerned, the plot indicates an alignment at $\hbar\omega$ $\approx$ 0.95.
It is possible that this alignment stems from the neutrons, and is \textquotedblleft delayed\textquotedblright owing to the blocking 
effect following the existing neutron occupancy of the $g_{9/2}$ orbital. Or, it is also probable that the alignment ascribes to the breaking 
of a proton pair in the $fp$ orbitals. Interestingly, the shell model calculations indicate that the average $g_{9/2}$ neutron occupancy 
sharply changes 
from $\approx$ 0.40 in the lower (5/2$^-_1$, 9/2$^-_1$, 13/2$^-_1$) states of the 5/2$^-$ band
to $\approx$ 2.08 at the 17/2$^-_1$ level.
It is also noteworthy that the proton occupancy of the $p_{3/2}$, $f_{5/2}$ and $p_{1/2}$ orbitals considerably evolves from the 5/2$^-_1$, 9/2$^-_1$, 13/2$^-_1$ states
to that in the higher (17/2$^-_1$, 21/2$^-$) levels. 
These changing occupancies along the band could be deemed as a reflection of the possibility of both the neutron and the proton alignments in the structure. 
The same 5/2$^-$ band in $^{67}$Zn was reported only up to limited 
excitation \cite{Zha25} and also incorporate an alignment feature, albeit at lower ($\hbar\omega$ $\approx$ 0.7) frequency. However, the latter proposition is 
based on a rather weak 1469 keV transition, newly identified in the last work \cite{Zha25}, that could be subject to tentative placement and other uncertainties. \\

The shape/ deformations associated with the band structures could be ascertained in the calculations of the Total Routhian Surface (TRS) that have been pursued in this work.
The procedure is elucidated in Ref.~\cite{Muk09} and is based on the Nilsson–Strutinsky framework. The single particle energies are calculated using the Woods-Saxon potential \cite{Naz85,Naz90}.
The total Routhian energies are calculated for varying rotational frequency ($\hbar\omega$) on a three-dimensional deformation grid that is defined by the (shape) parameters 
($\beta_2$, $\gamma$, $\beta_4$). The resulting potential energy surfaces (for individual $\hbar\omega$), following the minimization of $\beta_4$, 
are represented as contour plots in the $\beta_2$-$\gamma$ plane. The value $\gamma$ = 0$^o$ corresponds to a prolate shape and $\gamma$ = -60$^o$ corresponds to 
oblate (Lund convention). For a given nuclear configuration and rotational frequency, the equilibrium deformation is determined
by the location of the minima in the corresponding potential energy contour plot. Fig.~10 illustrates the TRS plot corresponding to the bandhead of the 
5/2$^-$ band, based on $\nu f_{5/2}$ configuration. It is noted that the potential energy minimum is at quadrupole deformation, $\beta_2$ $\approx$ 0.15, 
but with a considerable degree of $\gamma$-softness around the minimum. That represents a moderately deformed shape, subject to fluctuations in the $\gamma$ degree 
of freedom, associated with the ground state band of the $^{65}$Zn nucleus. The moderate deformation and the $\gamma$-softness are commensurate with 
its transitional character as well as with the limited impact of the $f_{5/2}$ occupancy on nuclear deformation. \\   

\begin{figure}
\includegraphics[angle=0,scale=.27,trim=0.0cm 0.0cm 0.0cm 0.0cm,clip=true]{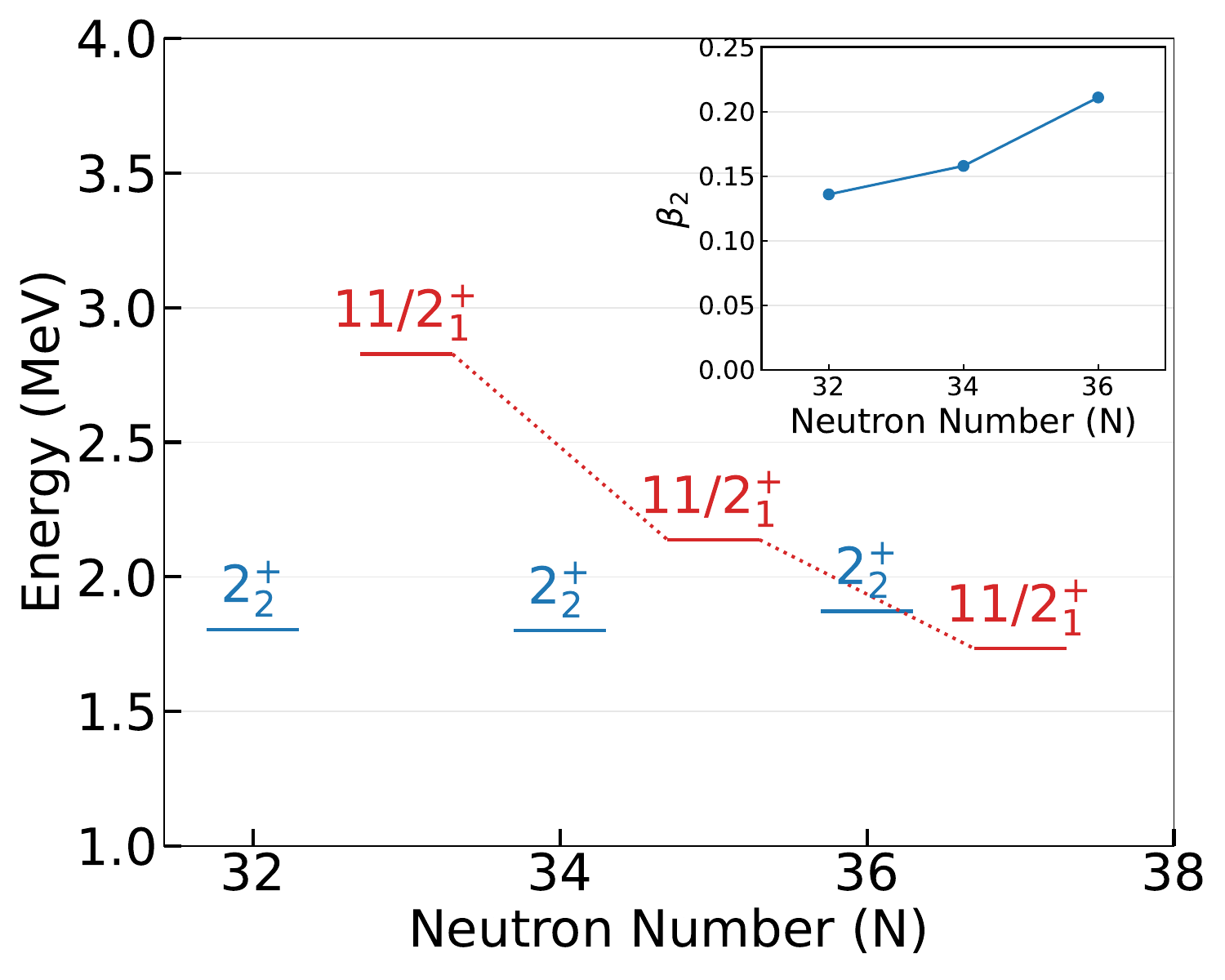}
\caption{\label{fig12}Excitation energy of the 2$_2^+$ state across the even-even isotopes of Zn and that of the 11/2$^+$ bandhead of the respective odd-neutron systems.
The $\beta$ values plotted in the inset are calculated from the respective TRS minima; the methodology of the calculations are described in the text.}
\end{figure}

\begin{figure}
\includegraphics[angle=0,scale=.27,trim=0.0cm 0.0cm 0.0cm 0.0cm,clip=true]{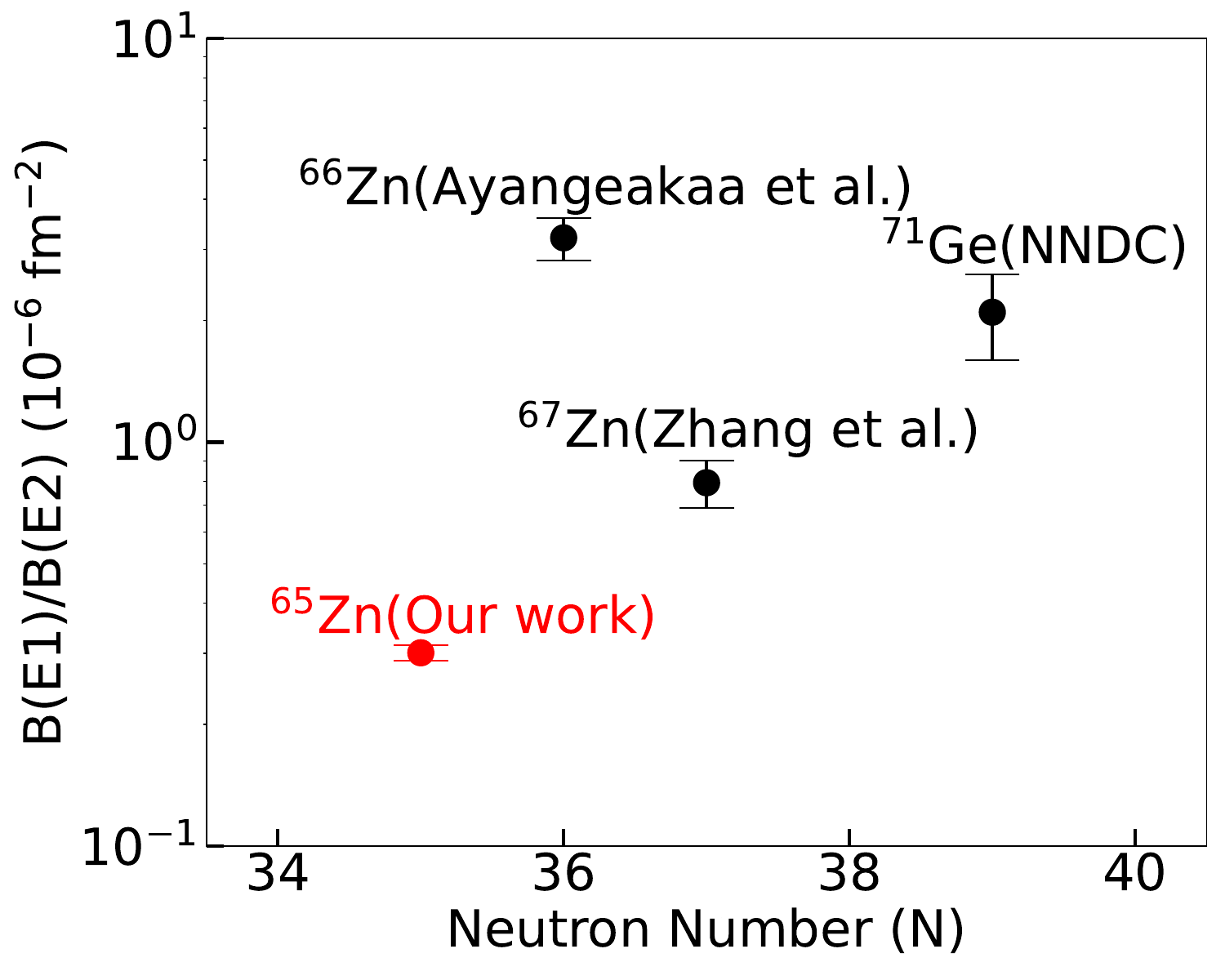}
\caption{\label{fig13}$B(E1)/B(E2)$ corresponding to the octupole excitations in the nuclei labeled in the plot. The $\gamma$-rays de-exciting the 19/2$^-$ state of the respective octupole sequence have 
been used for the purpose.}
\end{figure}

The level structures of both $^{65}$Zn and $^{67}$Zn isotopes highlight that the excitations principally proceed through the 9/2$^+$ band, based on the 
$\nu g_{9/2}$ configuration. It is noteworthy that the MOI characteristics of the band, illustrated in Fig.~9(a), are very similar in the two isotopes
and hint at similar deformations/ shapes associated with the respective sequence. The nature and the position of the alignments, plotted in Fig.~9(b),
are also overlapping for the two nuclei and could presumably be ascribed to a neutron pair. The shell model calculations, however, do not 
bring out any change in the neutron occupancy (Table~2) of the $g_{9/2}$ orbital throughout the sequence; the corresponding calculated (Table~2) energies are considerably deviant, 
particularly at the higher (21/2$^+$, 25/2$^+$) excitations. It is further interesting that the 
intensity of the $\gamma$-ray transitions forks into multiple branches above the 17/2$^+$ state of the $g_{9/2}$ band in both $^{65}$Zn and $^{67}$Zn nuclei. 
That might be indicative of competing mechanisms for generation of angular momentum that translate into multiple structures in the corresponding excitation regime. 
Such possibilities also emerge in the TRS plot of the $g_{9/2}$ band, at its bandhead and at higher excitation, as illustrated in Fig.~11. The potential minimum at the 
lowest frequency corresponds to $\beta_2$ $\approx$ 0.23 and $\gamma$ $\approx$ 90$^o$ that represents a stable equilibrium deformation. At higher frequencies ($\hbar\omega$),
say $\hbar\omega$ = 0.6, the potential energy surface exhibit multiple minima; one at $\beta_2$ $\approx$ 0.21 and a shallow one at $\beta_2$ $\approx$ 0.36. 
It is interesting to note that this frequency also overlaps with the one corresponding to the 17/2$^+$ state of the band, beyond which the excitation branches into multiple pathways.  
The appearance of competing minima in the potential energy surface at higher excitations could be commensurately suggestive of the coexistence of
different shape configurations and points to an increased softness of the potential energy surface with rotation. \\

The B3 sequence is based on the 11/2$^+$ bandhead at $E_x$ = 2137-keV. Most of the transitions in this band were previously \cite{Muk01} known. The $E_x$ versus $I(I+1)$ plot of the
sequence (Fig.~9(a)) closely follows that of the 9/2$^+$ band, particularly before the alignment in the latter; the MOI values (6.19 and 5.97 $\hbar^2/MeV$) of the two bands 
are rather close over the corresponding range of spins. The overlap strongly suggests that the 11/2$^+$ sequence results from the coupling of the 
anharmonic ($\beta$) vibration of the core to the structure based on the 9/2$^+$. It is interesting to track the excitation energy of the 11/2$^+$ state 
across the odd-A isotopic chain of Zn, with reference to the known $\beta$-vibrational 2$_2^+$ state \cite{Neu77} of the corresponding even-even core and keeping the evolving $\beta$-value
in perspective. That is illustrated in Fig.~12 wherein it is noted that the energy of the vibrational 2$_2^+$ state remains almost constant for the even-even isotopes. 
However, the steady decrease in the excitation energy of the 11/2$^+$ level, with increasing neutron number and $\beta$-deformation (illustrated in the inset of the plot), 
points to increasingly favored coupling between the vibrational and the rotational degrees of freedom across the chain. \\

The B4 sequence in the present level scheme of $^{65}$Zn is built on the 15/2$^-$ state and qualitatively resembles the similar structure in the neighboring $^{67}$Zn \cite{Zha25}.
The sequence here, just like that in $^{67}$Zn, exhibit regular rotational pattern and is connected to the positive parity 9/2$^+$ band by E1 transitions. Some 
of the latter have been newly identified in this work. As has been highlighted \cite{Zha25} for the $^{67}$Zn, the excitation energy of the 15/2$^-$ bandhead, with respect to the 
9/2$^+$ level, should trace the variation of energy of the 3$^-$ octupole excitation of the even-even Zn core. As illustrated in Fig.~7, the energy of the 15/2$^-$ state 
in $^{65}$Zn does fit in this trend, and so does the other states (19/2$^-$, 23/2$^-$, 27/2$^-$) of the sequence. The strength ($B(E1)$) of the interband E1 transitions vis-a-vis 
that ($B(E2)$) of the intra-band E2 ones, however, does not indicate a strong octupole correlation in $^{65}$Zn in comparison with the $^{67}$Zn isotope. 
Indeed the minimum (and thus most stable) octupole excitation of the even-even Zn core corresponds to $N = 38$. Thus it is valid to state that the behavior 
of the 15/2$^-$ sequence in $^{65}$Zn is, at best, a representation of the developing octupole collectivity/ correlation across the isotopic chain. As a quantification, the $B(E1)/B(E2)$ has been calculated 
for the de-excitations of the 19/2$^-$ state of the band, and compared with the equivalent values in the neighboring Zn isotopes and $^{71}$Ge. The results are illustrated in Fig.~13.   
The $B(E1, 1281-keV)/B(E2, 858-keV)$ for the $^{65}$Zn, as extracted in this analysis, is $\approx$ 0.3$\times$10$^{-6}$ fm$^{2}$ and is smaller than the same in $^{66,67}$Zn and in $^{71}$Ge. 
The latter has been evidenced \cite{Wan22} as an example of enhanced octupole correlation around $N \approx 40$. \\

\section{Conclusion}

The level structure of the $^{65}$Zn nucleus was investigated following its production in $\alpha$-beam induced fusion-evaporation reaction on $^{63}$Cu.
The $\gamma$-rays were detected using an array of Compton suppressed HPGe clover detectors. The conventional methodologies of experiment and data analysis were
pursued to establish the excitation scheme of the nucleus. The level structure has been interpreted both in the framework of large basis shell model calculations as well as 
in the context of the rotational characteristics of the observed bands. The data also reveals the footprints of developing octupole collectivity in the $fpg$ model space 
outside the $^{56}$Ni-core. The perspectives emerging from the study validate a transitional character of the $^{65}$Zn nucleus that 
exhibit discernible traits of both single particle excitations as well as collectivity underlying its level scheme.

\section*{Acknowledgments}

This work is partially supported by the Department of Science
and Technology, Government of India (No. IR/S2/PF-03/2003-II). SK acknowledges University Grants Commission, Government of India for fellowship under the NETJRF scheme. 
AD acknowledges Department of Science and Technology, Government of India for fellowship support under the INSPIRE scheme.

\bibliography{65Zn_rr}

\end {document}